\documentclass[preprint]{aastex631}

\newcommand{\new}{}

\usepackage{amsmath}
\usepackage{multirow}

\usepackage{graphicx}
\graphicspath{{Figures/}}

\begin{document}

\title{Exploring pulsar timing precision: A comparative study of polarization calibration methods for NANOGrav data from the  Green Bank Telescope}

\correspondingauthor{Lankeswar Dey}
\email{lankeswar.dey@nanograv.org}

\author[0000-0002-2554-0674]{Lankeswar Dey}
\affiliation{Department of Physics and Astronomy, West Virginia University, P.O. Box 6315, Morgantown, WV 26506, USA}
\affiliation{Center for Gravitational Waves and Cosmology, West Virginia University, Chestnut Ridge Research Building, Morgantown, WV 26505, USA}
\author[0000-0001-7697-7422]{Maura A. McLaughlin}
\affiliation{Department of Physics and Astronomy, West Virginia University, P.O. Box 6315, Morgantown, WV 26506, USA}
\affiliation{Center for Gravitational Waves and Cosmology, West Virginia University, Chestnut Ridge Research Building, Morgantown, WV 26505, USA}
\author[0000-0001-9678-0299]{Haley M. Wahl}
\affiliation{Department of Physics and Astronomy, West Virginia University, P.O. Box 6315, Morgantown, WV 26506, USA}
\affiliation{Center for Gravitational Waves and Cosmology, West Virginia University, Chestnut Ridge Research Building, Morgantown, WV 26505, USA}
\author[0000-0002-6664-965X]{Paul B. Demorest}
\affiliation{National Radio Astronomy Observatory, 1003 Lopezville Rd., Socorro, NM 87801, USA}
\author{Zaven Arzoumanian}
\affiliation{X-Ray Astrophysics Laboratory, NASA Goddard Space Flight Center, Code 662, Greenbelt, MD 20771, USA}
\author[0000-0003-4046-884X]{Harsha Blumer}
\affiliation{Department of Physics and Astronomy, West Virginia University, P.O. Box 6315, Morgantown, WV 26506, USA}
\affiliation{Center for Gravitational Waves and Cosmology, West Virginia University, Chestnut Ridge Research Building, Morgantown, WV 26505, USA}
\author[0000-0003-3053-6538]{Paul R. Brook}
\affiliation{Institute for Gravitational Wave Astronomy and School of Physics and Astronomy, University of Birmingham, Edgbaston, Birmingham B15 2TT, UK}
\author[0000-0003-4052-7838]{Sarah Burke-Spolaor}
\altaffiliation{Sloan Fellow}
\affiliation{Department of Physics and Astronomy, West Virginia University, P.O. Box 6315, Morgantown, WV 26506, USA}
\affiliation{Center for Gravitational Waves and Cosmology, West Virginia University, Chestnut Ridge Research Building, Morgantown, WV 26505, USA}
\author[0000-0002-6039-692X]{H. Thankful Cromartie}
\affiliation{National Research Council Postdoctoral Associate, National Academy of Sciences, Washington, DC 20001, USA resident at Naval Research Laboratory, Washington, DC 20375, USA}
\author[0000-0002-2185-1790]{Megan E. DeCesar}
\affiliation{Department of Physics and Astronomy, George Mason University, Fairfax, VA 22030}
\author[0000-0001-8885-6388]{Timothy Dolch}
\affiliation{Department of Physics, Hillsdale College, 33 E. College Street, Hillsdale, MI 49242, USA}
\affiliation{Eureka Scientific, 2452 Delmer Street, Suite 100, Oakland, CA 94602-3017, USA}
\author{Justin A. Ellis}
\altaffiliation{Infinia ML, 202 Rigsbee Avenue, Durham NC, 27701, USA}
\author[0000-0002-2223-1235]{Robert D. Ferdman}
\affiliation{School of Chemistry, University of East Anglia, Norwich, NR4 7TJ, United Kingdom}
\author[0000-0001-7828-7708]{Elizabeth C. Ferrara}
\affiliation{Department of Astronomy, University of Maryland, College Park, MD 20742, USA}
\affiliation{Center for Research and Exploration in Space Science and Technology, NASA/GSFC, Greenbelt, MD 20771}
\affiliation{NASA Goddard Space Flight Center, Greenbelt, MD 20771, USA}
\author[0000-0001-5645-5336]{William Fiore}
\affiliation{Department of Physics and Astronomy, West Virginia University, P.O. Box 6315, Morgantown, WV 26506, USA}
\affiliation{Center for Gravitational Waves and Cosmology, West Virginia University, Chestnut Ridge Research Building, Morgantown, WV 26505, USA}
\author[0000-0001-8384-5049]{Emmanuel Fonseca}
\affiliation{Department of Physics and Astronomy, West Virginia University, P.O. Box 6315, Morgantown, WV 26506, USA}
\affiliation{Center for Gravitational Waves and Cosmology, West Virginia University, Chestnut Ridge Research Building, Morgantown, WV 26505, USA}
\author[0000-0001-6166-9646]{Nate Garver-Daniels}
\affiliation{Department of Physics and Astronomy, West Virginia University, P.O. Box 6315, Morgantown, WV 26506, USA}
\affiliation{Center for Gravitational Waves and Cosmology, West Virginia University, Chestnut Ridge Research Building, Morgantown, WV 26505, USA}
\author[0000-0001-8158-683X]{Peter A. Gentile}
\affiliation{Department of Physics and Astronomy, West Virginia University, P.O. Box 6315, Morgantown, WV 26506, USA}
\affiliation{Center for Gravitational Waves and Cosmology, West Virginia University, Chestnut Ridge Research Building, Morgantown, WV 26505, USA}
\author[0000-0003-4090-9780]{Joseph Glaser}
\affiliation{Department of Physics and Astronomy, West Virginia University, P.O. Box 6315, Morgantown, WV 26506, USA}
\affiliation{Center for Gravitational Waves and Cosmology, West Virginia University, Chestnut Ridge Research Building, Morgantown, WV 26505, USA}
\author[0000-0003-1884-348X]{Deborah C. Good}
\affiliation{Department of Physics and Astronomy, University of Montana, 32 Campus Drive, Missoula, MT 59812}
\author[0000-0003-1082-2342]{Ross J. Jennings}
\altaffiliation{NANOGrav Physics Frontiers Center Postdoctoral Fellow}
\affiliation{Department of Physics and Astronomy, West Virginia University, P.O. Box 6315, Morgantown, WV 26506, USA}
\affiliation{Center for Gravitational Waves and Cosmology, West Virginia University, Chestnut Ridge Research Building, Morgantown, WV 26505, USA}
\author[0000-0001-6607-3710]{Megan L. Jones}
\affiliation{Center for Gravitation, Cosmology and Astrophysics, Department of Physics, University of Wisconsin-Milwaukee,\\ P.O. Box 413, Milwaukee, WI 53201, USA}
\author[0000-0003-0721-651X]{Michael T. Lam}
\affiliation{SETI Institute, 339 N Bernardo Ave Suite 200, Mountain View, CA 94043, USA}
\affiliation{School of Physics and Astronomy, Rochester Institute of Technology, Rochester, NY 14623, USA}
\affiliation{Laboratory for Multiwavelength Astrophysics, Rochester Institute of Technology, Rochester, NY 14623, USA}
\author[0000-0003-1301-966X]{Duncan R. Lorimer}
\affiliation{Department of Physics and Astronomy, West Virginia University, P.O. Box 6315, Morgantown, WV 26506, USA}
\affiliation{Center for Gravitational Waves and Cosmology, West Virginia University, Chestnut Ridge Research Building, Morgantown, WV 26505, USA}
\author[0000-0001-5373-5914]{Jing Luo}
\altaffiliation{Deceased}
\affiliation{Department of Astronomy \& Astrophysics, University of Toronto, 50 Saint George Street, Toronto, ON M5S 3H4, Canada}
\author[0000-0001-5229-7430]{Ryan S. Lynch}
\affiliation{Green Bank Observatory, P.O. Box 2, Green Bank, WV 24944, USA}
\author[0000-0002-3616-5160]{Cherry Ng}
\affiliation{Dunlap Institute for Astronomy and Astrophysics, University of Toronto, 50 St. George St., Toronto, ON M5S 3H4, Canada}
\author[0000-0002-6709-2566]{David J. Nice}
\affiliation{Department of Physics, Lafayette College, Easton, PA 18042, USA}
\author[0000-0001-5465-2889]{Timothy T. Pennucci}
\affiliation{Institute of Physics and Astronomy, E\"{o}tv\"{o}s Lor\'{a}nd University, P\'{a}zm\'{a}ny P. s. 1/A, 1117 Budapest, Hungary}
\author[0000-0002-8826-1285]{Nihan S. Pol}
\affiliation{Department of Physics and Astronomy, Vanderbilt University, 2301 Vanderbilt Place, Nashville, TN 37235, USA}
\author[0000-0001-5799-9714]{Scott M. Ransom}
\affiliation{National Radio Astronomy Observatory, 520 Edgemont Road, Charlottesville, VA 22903, USA}
\author[0000-0002-6730-3298]{Ren\'{e}e Spiewak}
\affiliation{Jodrell Bank Centre for Astrophysics, University of Manchester, Manchester, M13 9PL, United Kingdom}
\author[0000-0001-9784-8670]{Ingrid H. Stairs}
\affiliation{Department of Physics and Astronomy, University of British Columbia, 6224 Agricultural Road, Vancouver, BC V6T 1Z1, Canada}
\author[0000-0002-7261-594X]{Kevin Stovall}
\affiliation{National Radio Astronomy Observatory, 1003 Lopezville Rd., Socorro, NM 87801, USA}
\author[0000-0002-1075-3837]{Joseph K. Swiggum}
\altaffiliation{NANOGrav Physics Frontiers Center Postdoctoral Fellow}
\affiliation{Department of Physics, Lafayette College, Easton, PA 18042, USA}





\begin{abstract}
Pulsar timing array experiments have recently uncovered evidence for a nanohertz gravitational wave background by precisely timing an ensemble of millisecond pulsars. 
The next significant milestones for these experiments include characterizing the detected background with greater precision, identifying its source(s), and detecting continuous gravitational waves from individual supermassive black hole binaries.
To achieve these objectives, generating accurate and precise times of arrival of pulses from pulsar observations is crucial. 
Incorrect polarization calibration of the observed pulsar profiles may introduce errors in the measured times of arrival.
Further, previous studies \citep[e.g.,][]{vanStraten2013, ManchesterHobbs+2013} have demonstrated that robust polarization calibration of pulsar profiles can reduce noise in the pulsar timing data and improve timing solutions.
In this paper, we investigate and compare the impact of different polarization calibration methods on pulsar timing precision using three distinct calibration techniques: the Ideal Feed Assumption (IFA), Measurement Equation Modeling (MEM), and Measurement Equation Template Matching (METM). 
Three NANOGrav pulsars---PSRs J1643$-$1224, J1744$-$1134, and J1909$-$3744---observed with the 800 MHz and 1.5 GHz receivers at the Green Bank Telescope (GBT) are utilized for our analysis. 
Our findings reveal that all three calibration methods enhance timing precision compared to scenarios where no polarization calibration is performed. 
Additionally, among the three calibration methods, the IFA approach generally provides the best results for timing analysis of pulsars observed with the GBT receiver system. 
We attribute the comparatively poorer performance of the MEM and METM methods to potential instabilities in the reference noise diode coupled to the receiver and temporal variations in the profile of the reference pulsar, respectively.
\end{abstract}

\keywords{Millisecond pulsars(1062) --- Pulsar timing method(1305) --- Astronomical techniques(1684)}


\section{Introduction} \label{sec:intro}

Pulsars are highly magnetized rapidly rotating neutron stars that emit beams of electromagnetic radiation along their magnetic axes. 
As the pulsar rotates, the beam(s) sweep(s) across the observer's line of sight and pulses are seen at regular intervals.
The remarkable rotational stability of pulsars, especially millisecond pulsars (MSPs), makes them invaluable tools in astrophysics. 
\new{MSPs can serve as very accurate celestial clocks and are used for testing theories of gravity, probing the interstellar medium (ISM), and detecting low-frequency gravitational waves (GWs)} \citep{LorimerKramer2004hpa}.
At the heart of these scientific endeavors employing MSPs lies the crucial concept of \textit{pulsar timing}.

In pulsar timing, the rotation of a pulsar is accurately tracked by measuring the times of arrival (TOAs) of its pulses and comparing them to the TOAs predicted from a pulsar timing model \new{\citep{LorimerKramer2004hpa, Hobbs+2006}}. 
Deviations of the observed TOAs from the predicted ones, i.e., the timing residuals, can reveal important information about the pulsar, its environment, and GW signals present between the Earth and the pulsar \new{\citep[][and references therein]{Taylor1979, Taylor1992, LorimerKramer2004hpa, Manchester2017}}.
GWs, tidal ripples in the fabric of spacetime, induce minute changes in the TOAs from the pulsars, and pulsar timing array (PTA) experiments observe an ensemble of MSPs to measure those changes in order to detect GWs. 
Recently, evidence for the presence of a low-frequency stochastic GW background (GWB) in their respective PTA data sets was reported independently by the North American Nanohertz Observatory for Gravitational Waves \citep[NANOGrav:][]{NG2023_15yr_GWB}, the European Pulsar Timing Array (EPTA) + Indian Pulsar Timing Array (InPTA) \citep{AntoniadisArumugam+2023c}, the Parkes Pulsar Timing Array \citep[PPTA:][]{ReardonZic+2023}, and the Chinese Pulsar Timing Array \citep[CPTA:][]{XuChen+2023} collaborations.

The next significant milestones for PTAs to accomplish are characterizing the observed GWB more precisely, determining its source(s), and detecting continuous GWs from individual supermassive black hole binaries (SMBHBs).
The PTA sensitivity directly depends upon the precision and accuracy with which pulse arrival times can be estimated and therefore generating accurate and precise TOAs from pulsar observations is a critical aspect of the PTA experiments.
Observing pulsars using bigger and better telescopes with larger bandwidths and longer integration times is one of the possible ways to improve the precision of TOAs.
Better precision and accuracy in TOAs can also be achieved by developing improved methods of TOA estimation, radio frequency interference (RFI) mitigation, and instrumental calibration. 
\new{Additionally, inaccurate polarization calibration can distort  the observed pulsar polarization profiles and therefore introduce noise in measured TOAs \citep[see e.g.,][]{vanStraten2013,Foster2015, Guillemot2023, Rogers2024}.}
Hence, adopting a robust and accurate polarization calibration procedure to calibrate pulsar profiles holds the potential to elevate the precision and accuracy of measured TOAs, thereby facilitating the realization of the current scientific goals set by PTA collaborations.

Pulsars are among the most polarized of all known radio sources and \new{polarization measurements} can provide additional insights into the pulsar emission process and the medium through which the radiation propagates.
The polarization state of a pulsar signal can be described by the four Stokes parameters $I, Q, U$, and $V$, and can be represented by the Stokes vector \citep{Stokes1851}
\begin{equation}
    S = \begin{bmatrix}
  I \\
  Q \\
  U \\
  V \\
\end{bmatrix}\,,
\end{equation}
where $I$ is the total intensity, $Q$ and $U$ form the linear polarization $L = \sqrt{Q^2 + U^2}$, and $V$ represents the circular polarization intensity.
Using the International Astronomical Union’s (IAU’s) convention, right-handed circular polarization is positive and left-handed circular polarization is negative \citep{Stokes1851}.
As the radio waves from a pulsar travel through the ISM, they experience Faraday rotation, a frequency-dependent rotation of the polarization position angle caused by the Galactic magnetic field.
Interstellar Faraday rotation ($\beta$) can be given by
\begin{equation}
    \beta =RM \,\, \lambda^2 \,,
\end{equation}
where $\lambda$ is the wavelength of the radio waves and the overall strength of the effect is characterized by the rotation measure ($RM$).
The $RM$ depends on the interstellar magnetic field component ($B_{||}$) parallel to the line of sight and free-electron density ($n_e$) as (in cgs units)
\begin{equation}
    RM = \frac{e^3}{2\pi m_e^2 c^4} \int_{0}^{d} n_e(l) B_{||}(l)\, dl \,,
\end{equation}
where $e$ and $m_e$ are the charge and mass of electron, respectively, $c$ is the speed of light in vacuum, and $d$ is the distance to the pulsar.
Measurements of Faraday rotations in linearly polarized pulsars are used to study the ISM and the large-scale Galactic magnetic field in the Milky Way \citep{Han2018, Sobey2019}.

When a pulsar is observed with a radio telescope, the processes of reception and detection of the signal introduce instrumental artifacts that make the measured Stokes vector $S_{\text{m}}$ differ from the intrinsic Stokes vector $S_{\text{i}}$.
We can relate $S_{\text{m}}$ to $S_{\text{i}}$ with the Mueller matrix $M$ such that
\begin{equation}
    S_{\text{m}} = M\,S_{\text{i}}\,,
    \label{eq:mueller}
\end{equation}
where $M$ depends on differential gain and differential phase of the receiver, and ellipticity and non-orthogonality of the feeds \citep{Heiles+2001}.
During the polarization calibration, we determine $M$ by calibrating the observing system and thereafter solve Equation~\eqref{eq:mueller} to obtain the true Stokes vector $S_{\text{i}}$ from the observed $S_{\text{m}}$.
However, there are different methods for performing the pulsar polarization calibration proposed in the literature.

In the NANOGrav data releases, the pulsar profiles are calibrated using \new{the} Ideal Feed Assumption (IFA) which assumes the feeds to be perfectly linear and orthogonal.
However, in reality, this assumption may not be valid, and therefore using IFA-calibrated pulsar profiles to generate TOAs can introduce systematic errors due to polarization miscalibration.
Better polarization calibration methods, where a full polarimetric response (PR) of the observing system is used to calibrate the pulsar profiles, have been proposed to overcome the shortcomings of the IFA approach.
\cite{vanStraten2004} developed \new{the} Measurement Equation Modeling (MEM) that uses a pulsar with strong linear polarization observed over a wide range of parallactic angles to estimate the full PR of the observing system.
Additionally, \new{the} Measurement Equation Template Matching (METM) was introduced by \cite{vanStraten2013} as a polarization calibration method that matches multiple observations of a reference pulsar to a well-calibrated template profile of that pulsar to generate precise PR solutions at different epochs.
In both the MEM and METM methods, the calculated receiver solutions are used to perform polarization calibration of the observed pulsar profiles, and then accurate and precise TOAs can be generated from the calibrated profiles.

The effects of these robust polarization calibration techniques on the accuracy of the timing analysis for different pulsars have been explored in many studies.
\cite{ManchesterHobbs+2013} have found that, for 9 of the 20 pulsars observed with the Parkes radio telescope at 20-cm band, MEM-calibrated profiles provided timing residuals with reduced root-mean-square (RMS) and chi-squared values compared to uncalibrated profiles.
However, for \new{the} rest of the pulsars, the MEM calibration made little difference to the reduced chi-squared of the timing solution.
\cite{vanStraten2013} compared the timing accuracy of PSR J1022+1001 for different combinations of polarization calibration and TOA generation methods using the Parkes telescope data.
Two different methods for TOA generation were used: matrix template matching (MTM) which uses the full polarization profiles for generating TOAs \citep{vanStraten2006}, and standard total intensity (STI) which uses only the total intensity profile.
They found that calibration with the METM method reduces both the standard deviation of the arrival time residuals and the reduced chi-squared of the model fit compared to IFA calibration, for both the STI and MTM methods of TOA generation.
\new{\cite{Guillemot2023} also used improved polarization calibration methods on Nan{\c{c}}ay Radio Telescope data and found that TOAs generated from MEM-calibrated data with the STI-method have lower RMS and reduced chi-squared value compared to those generated using IFA calibration. They also showed that using the MTM-method for TOA generation further improved the timing quality.}
A detailed analysis of the effects of different polarization calibration methods on pulsar TOAs was also conducted by \cite{Rogers2020} for five pulsars observed with the Parkes radio telescope.
The study found that METM polarization calibration combined with MTM for TOA generation gives better timing accuracy compared to traditional IFA calibration followed by STI for TOA generation for all five pulsars.

Furthermore, \cite{Gentile2018} performed METM polarization calibration on a subset of NANOGrav data observed with the Arecibo Telescope to obtain some of the most sensitive polarimetric MSP profiles. 
This was repeated for a subset of Green Bank Telescope (GBT) pulsar data in \cite{Wahl2022} where they used MEM calibration method for the polarization calibration.
However, a detailed timing analysis with those calibrated profiles \new{has} not been done yet (see \citeauthor{Haley_Thesis}~\citeyear{Haley_Thesis} for a precursor work).
These results motivated us to explore the effects of different polarization calibration methods on the timing analysis for a subset of NANOGrav pulsars observed at the GBT.

In this paper, we present the timing analysis of three pulsars: PSRs J1643$-$1224, J1744$-$1134, and J1909$-$3744, with different polarization calibration methods.
We have used three different methods for performing polarization calibration of the pulsar profiles: (i) IFA, (ii) MEM, and (iii) MEM+METM, and TOAs were generated from the calibrated profiles using the STI.
The sets of TOAs generated for different polarization calibration methods are then individually used to perform timing analysis to compare the influence of different polarization calibrations on timing analysis for these pulsars.

The paper is structured as follows. 
In Section~\ref{sec:data}, we briefly describe the data we used in our paper.
Different polarization calibration methods are discussed in Section~\ref{sec:methods} along with a brief overview of the timing analysis procedure used in this paper.
Section~\ref{sec:results} contains the results of our analyses and the results are summarized and discussed in detail in Section~\ref{sec:discussion}.

\section{Data} \label{sec:data}

In this paper, we focus on three pulsars observed with the 100-meter Green Bank Telescope in the NANOGrav program: PSRs J1643$-$1224, J1744$-$1134, and J1909$-$3744.
We only use a subset of the data taken with the Green Bank Ultimate Pulsar Processing Instrument \citep[GUPPI;][]{DuPlain2008} backend system at both 820 MHz (with Rcvr\_800) and 1500 MHz (with Rcvr1\_2) frequencies with bandwidths of 200 MHz and 800 MHz, respectively.
PSR J1643$-$1224 is a bright pulsar with average flux densities of 12.9 and 4.7 mJy at 820 MHz and 1500 MHz frequencies, respectively, and has a moderate polarization fraction ($\sim 21\%$ at both 820 and 1500 MHz frequencies).
This pulsar was examined in \cite{Rogers2020}, which allows a direct comparison of our results.
PSR J1744$-$1134 has average flux densities of 6.2/2.6 mJy at 820/1500 MHz frequencies, is highly polarized (polarization fractions at 820/1500 MHz are $\sim 78\%/88\%$), \new{and was studied in \cite{Guillemot2023}}.
PSR J1909$-$3744, one of the best pulsars in NANOGrav, also has a fairly high polarization fraction (the polarization fractions at 820/1500 MHz are $\sim 51\%/45\%$ and the average flux densities are 3.6/1.3 mJy).

Although the GUPPI data acquisition instrument was used at GBT from 2010 March to 2020 April for NANOGrav observations, we only use the data from 2010 March (MJD 55265) to 2014 March (MJD 56739) in this paper.
This is because of a technical problem, which arose in 2014 March, that made the time alignment of the digitizers for the X and Y polarization of the telescope signals unstable, thus corrupting the polarization cross products.
Therefore, it is impossible to recover the correct full Stokes parameters from the data taken after March 2014.
\new{However, the power in the two individual polarizations (and thus Stokes $I$ and $Q$) remained unaffected, allowing well-calibrated total intensity profiles to be produced using the IFA calibration method.
Consequently, this instability should not impact the timing data after March 2014 in NANOGrav data releases that utilize IFA polarization calibration and total intensity profiles to generate the TOAs \citep[see][for more details]{Wahl2022}.}

The pulsars were observed with an approximately monthly cadence and the data were coherently de-dispersed, with frequency resolution of $\sim 1.56$ MHz.
The resulting time series were folded in real time using a nominal pulsar timing model to obtain folded pulse profiles as functions of time, radio frequency, and polarization.
The folded profiles have 2048 phase bins and subintegrations of 10 seconds.
We perform polarization calibrations on these folded profiles to get accurately calibrated pulsar profiles and thereafter use them to generate TOAs for timing analysis.
In addition, we used three long-track \new{($\sim$3.5 hours)} observations of PSRs B1929+10 \citep[on MJDs 56244, 56419, and 56608;][]{Kramer2021} at 820 MHz and one long-track observation of J1022+1001 (MJD 55671) at 1500 MHz with the GUPPI backend system to generate full PRs of the observing system for the MEM calibration method \new{(discussed in more detail in the next section)}.
Additionally, observations of PSR B1937+21 at GBT during MJD $55265-56739$ are used as reference pulsar profiles for performing the METM polarization calibration.

\section{Methods} \label{sec:methods}

In this section, we discuss various methods employed in this paper.
Different polarization calibration methods, namely the IFA, MEM, and METM, are described in detail in Subsection~\ref{subsec:polcals}.
Subsection~\ref{subsec:toa_timing} outlines our methods for TOA generation and timing analysis.

\subsection{Polarization Calibration Methods}
\label{subsec:polcals}

As discussed in Section~\ref{sec:intro}, the measured polarization profiles or the Stokes vector ($S_{\text{m}}$) of a pulsar observed with a radio telescope differs from the intrinsic one ($S_{\text{i}}$) due to instrumental effects.
The measured and intrinsic Stokes vectors are related by the Mueller matrix $M$ (see \ref{eq:mueller}) that characterizes the PR of the telescope.
The Mueller matrix for a dual linear feeds can be written in the form \citep[following][]{LorimerKramer2004hpa, Heiles+2001, Gentile2018}
\begin{equation}
M = \begin{bmatrix}
1 & E & A + EC & B + ED\\
E & 1 & AE + C & BE + D\\
AF - BG & DG - CF & F & -H\\
AG + BF & -CG - DF & G & F\\
\end{bmatrix}\,,
\end{equation}
where
\begin{eqnarray*}
A &=& e_1 \cos{\phi_1} + e_2\cos{\phi_2}\,,\\
B &=& e_1 \sin{\phi_1} + e_2\sin{\phi_2}\,,\\
C &=& e_1 \cos{\phi_1} - e_2\cos{\phi_2}\,,\\
D &=& e_1 \sin{\phi_1} - e_2\sin{\phi_2}\,,\\
E &=& \gamma/2\,,\\
F &=& \cos{\phi}\,,\\
H &=& \sin{\phi}\,,
\end{eqnarray*}
\new{and} $e_1$ and $e_2$ represent the magnitude of the cross-coupling of the two respective feeds, $\phi_1$ and $\phi_2$ represent the phase of this cross-coupling, $\gamma$ and $\phi$ represent the differential gain and phase of the receiver system, respectively.
By calibrating the observing system, we can determine $M$ and solve Equation~\eqref{eq:mueller} for the true Stokes vector $S_{\text{i}}$.
In different polarization calibration methods, the instrumental PR is calculated differently and therefore the calibrated profiles (i.e., the $S_{\text{i}}$'s generated by solving Equation~\eqref{eq:mueller}) could be different.
We now describe the different polarization calibration methods used in this paper.

\subsubsection{Ideal Feed Assumption}

\begin{figure}
\gridline{\fig{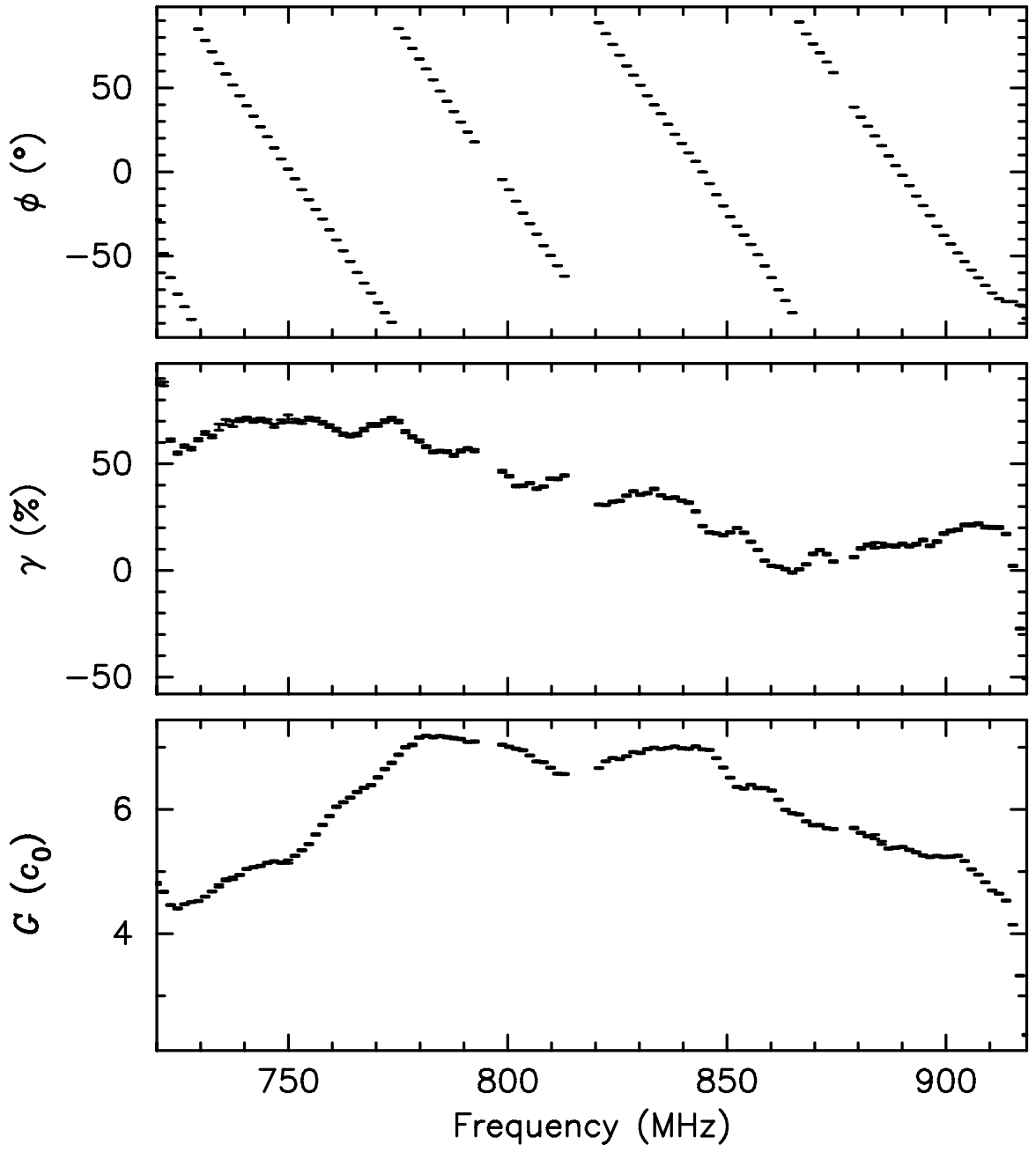}{0.3\textwidth}{(a) IFA (MJD 56614)}
          \fig{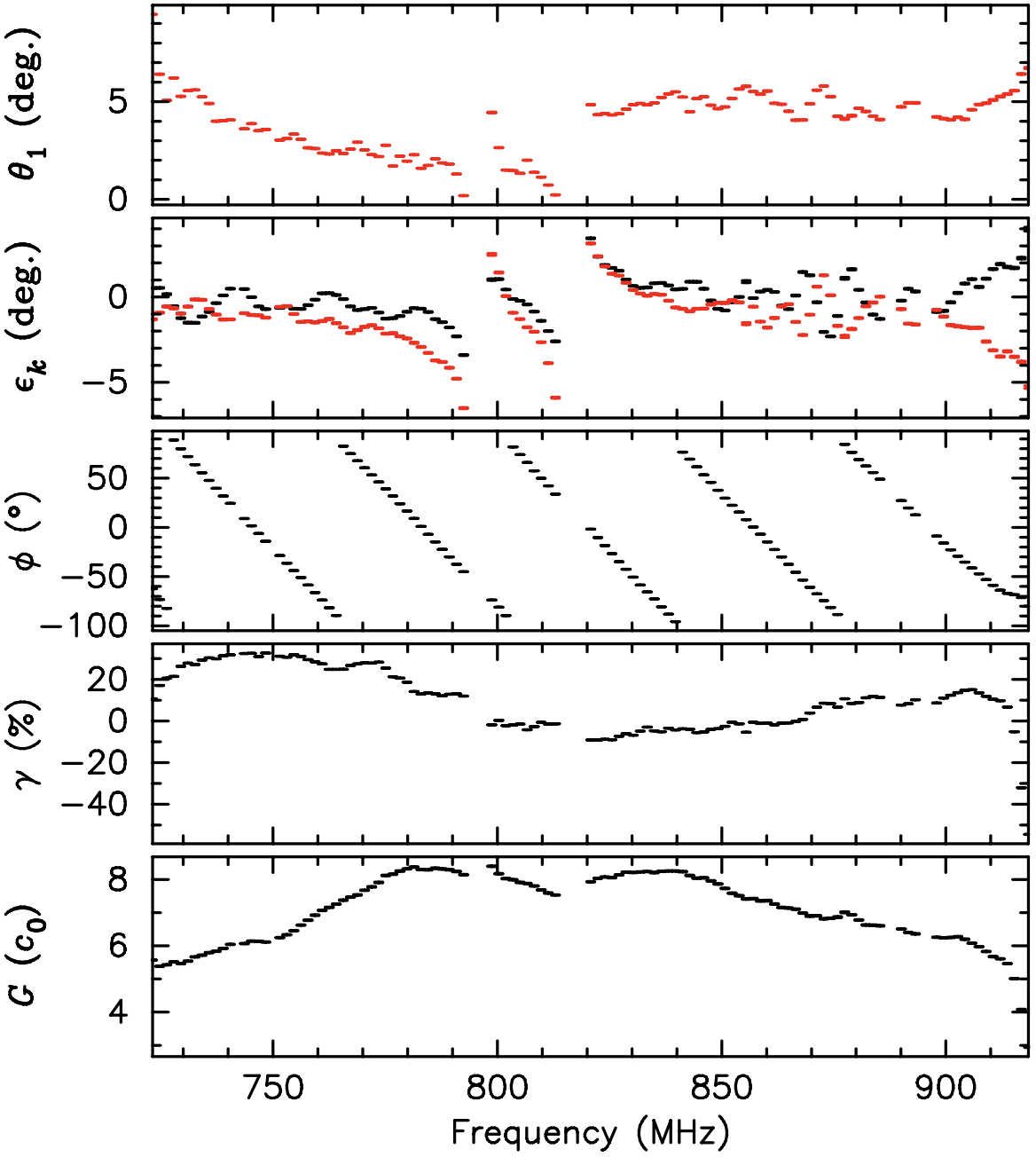}{0.3\textwidth}{(b) MEM (MJD 56244)}
          \fig{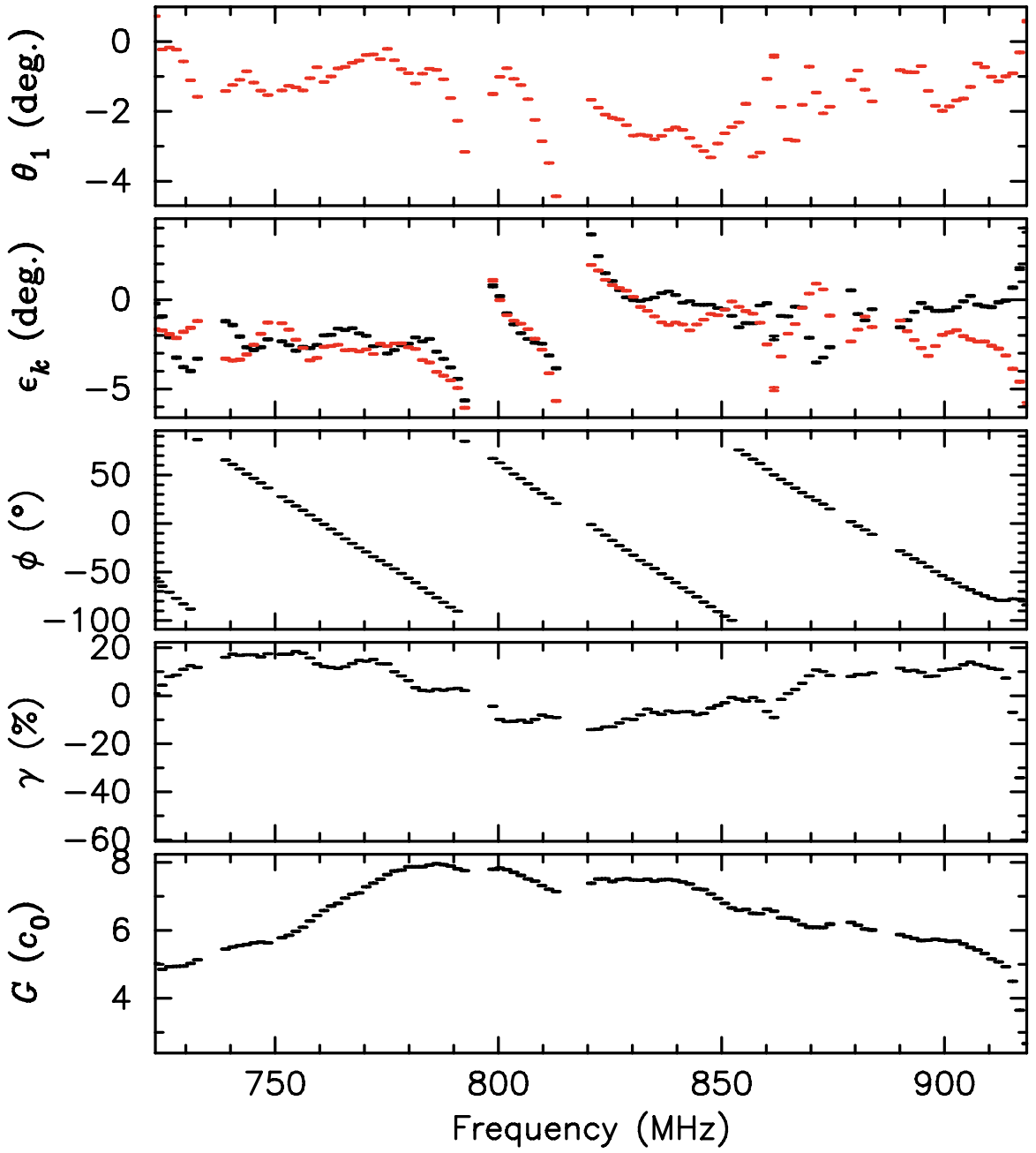}{0.3\textwidth}{(c) MEM (MJD 56419)}}
\gridline{\fig{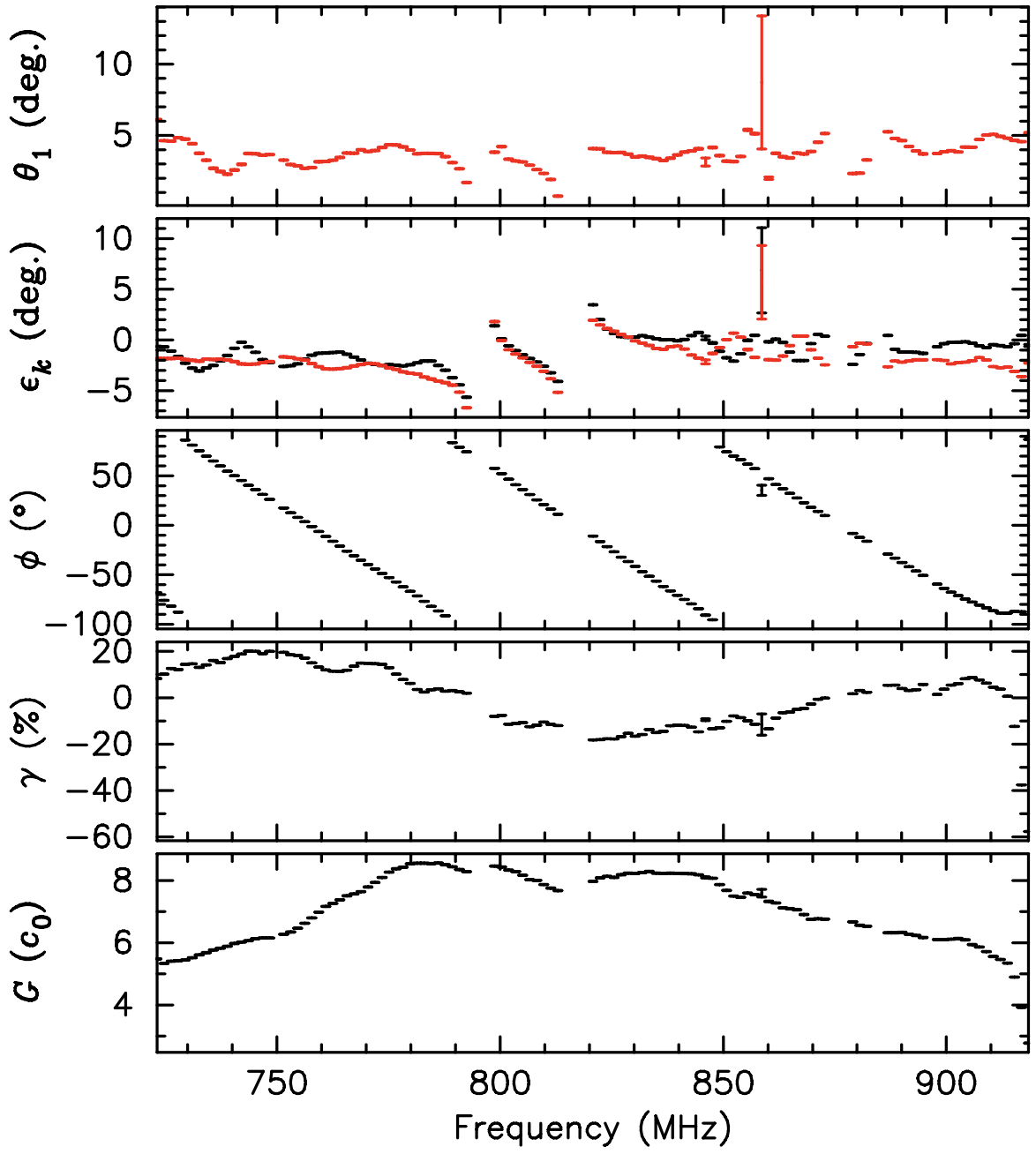}{0.3\textwidth}{(d) MEM (MJD 56608)}
          \fig{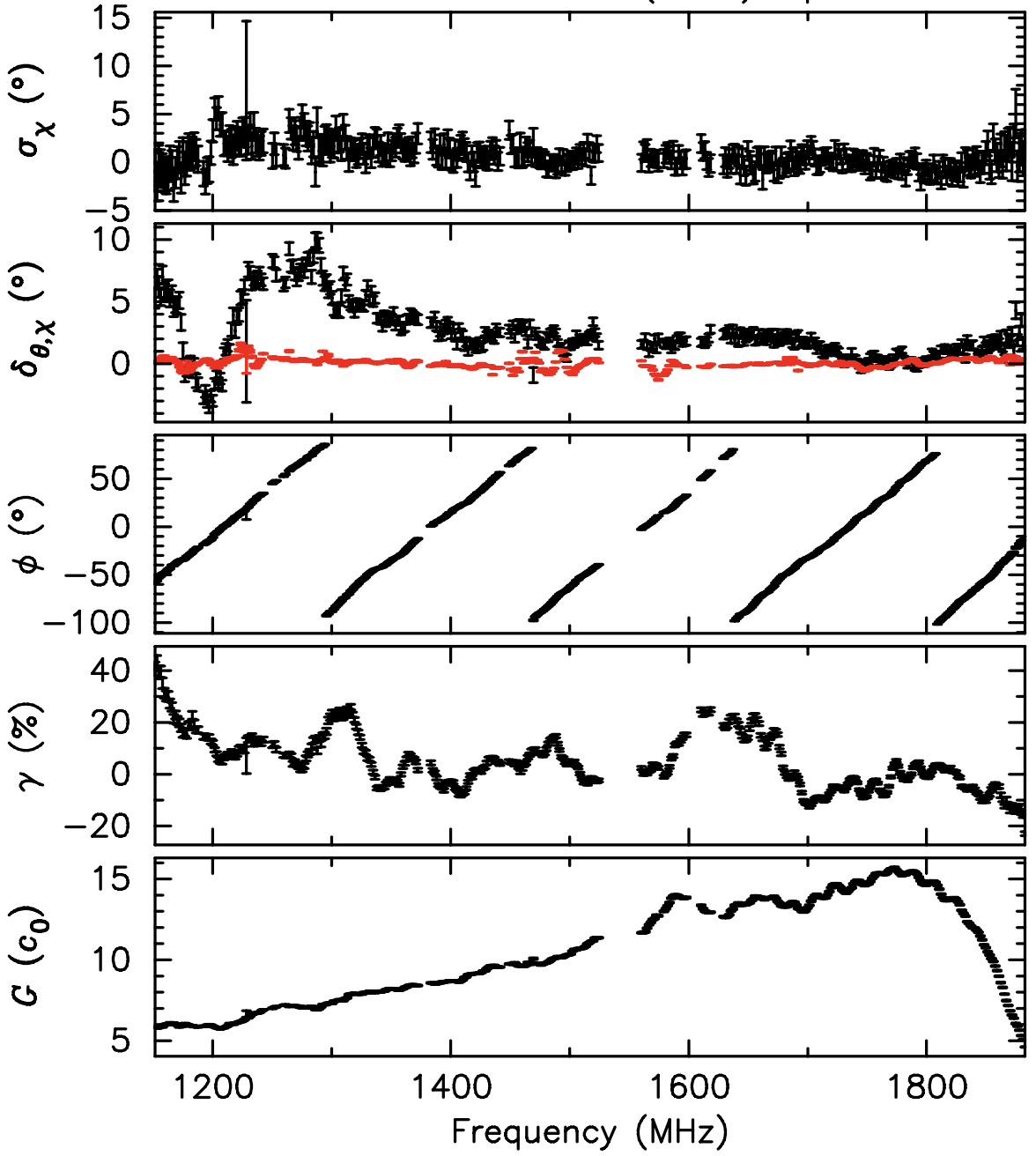}{0.3\textwidth}{(e) MEM (MJD 55671)}
          \fig{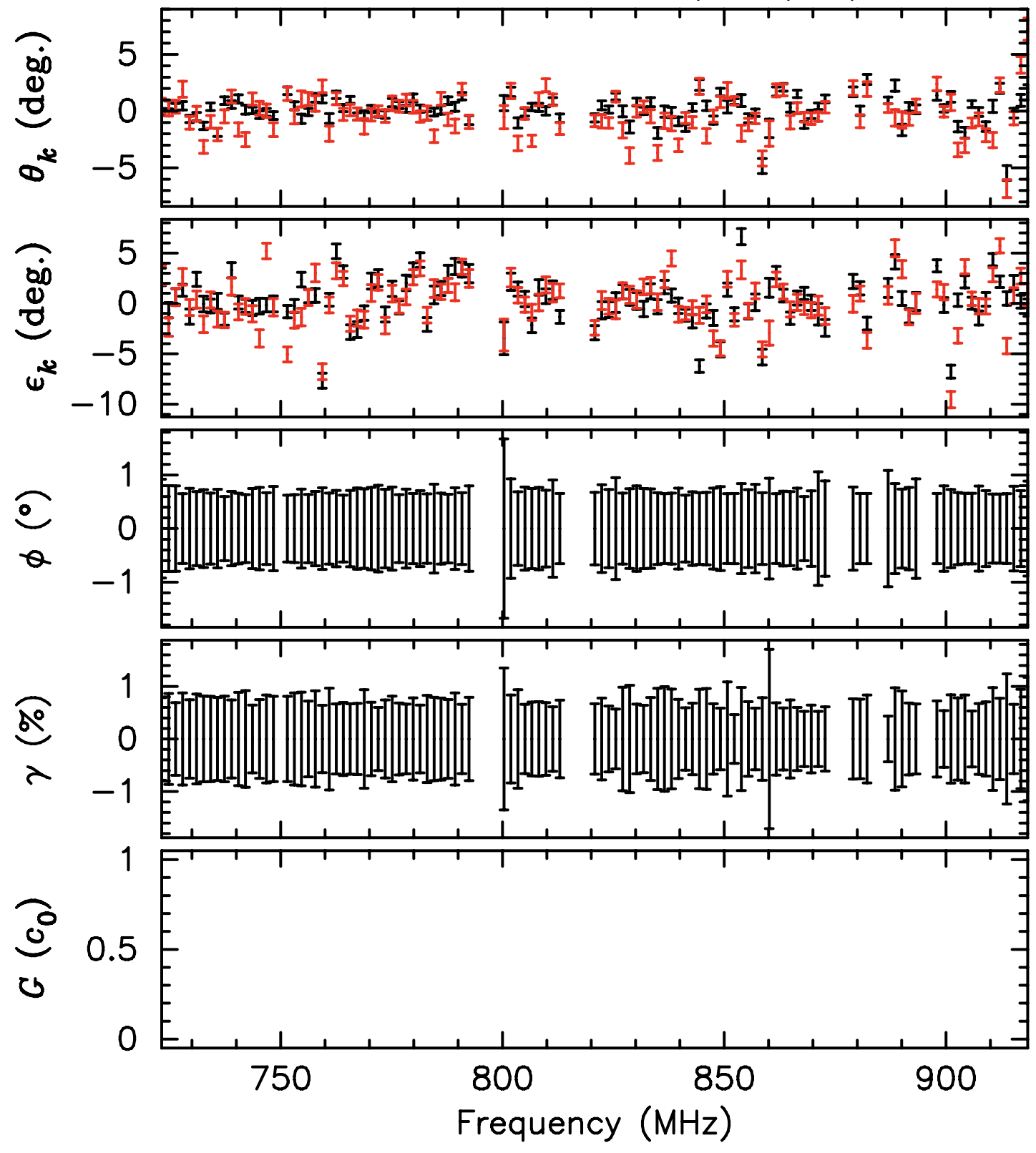}{0.3\textwidth}{(f) METM (MJD 56614)}}
\caption{Example polarimetric response solutions used for different polarization calibration methods to calibrate the observed pulsar profiles. 
In all the panels $G$, $\gamma$, and $\phi$ represent the absolute gain, differential gain, and differential phase of the observing system, respectively.
The absolute gain G is specified in units of the square root of the reference flux density ($c_0$).
Panel (a): IFA PR solution obtained from the reference noise diode observation for PSR J1744$-$1134 on MJD 56614.
\new{Panels (b), (c), and (d): MEM PR solutions at 800 MHz calculated from long track ($\sim 3.5$ hours) observations  of PSR B1929$+$10 on MJDs 56244, 56419, and 56608, respectively.}
Here, $\theta_1$ represents the orientation of receptor 1 with respect to receptor 0, and $\epsilon_k$ is the ellipticity angles of the two receptors (denoted by black and red points) \citep[see][for details]{vanStraten2004}.
\new{Panel (e): MEM PR solution for Rcvr1\_2, where $\delta_{\theta, \chi}$ and $\sigma_{\chi}$ represent the quantities as defined by equation~17 of \cite{Britton2000}.}
Panel (f): METM-generated PR correction, calculated using \new{a PSR~B1937$+$21 observation} on MJD 56614, to the MEM-generated PR on MJD 56608.
Here, $\theta_k$ and $\epsilon_k$ represent the orientations and ellipticities of the two receptors (black and red points), respectively.
Note that the absolute gain is not calculated for the METM-generated PR correction as we used the total invariant interval to normalize the Stokes parameters.
See Section~\ref{subsec:polcals} for more details.}
\label{fig:PRs}
\end{figure}

\new{The} Ideal Feed Assumption (IFA) is the most basic polarization calibration method and has been employed for all NANOGrav data releases \citep{StinebringCordes1984, NG2015_9yr_timing, NG2021_12yr_timing, NG2023_15yr_timing}.
As the signals from the two feeds during the observation pass through slightly different amplifier chains, it introduces different gains and phases on the signal. 
In this method, observation of a reference source is used to determine the differential gain ($\gamma$) and phase ($\phi$) of the receiver system.
\new{However, the magnitude ($e_i$) and phase ($\phi_i$) of the cross-coupling of the feeds are assumed to be zero in this method.}
For NANOGrav observations, the reference source is an artificial noise diode coupled to the receptors, emitting a square wave with a 50\% duty cycle and a period of 40 ms.
The noise diode is expected to be 100\% linearly polarized and should illuminate both receptors equally and in phase.
This noise diode is observed before each pulsar observation and the observations are used to determine the complex gains of the instrumental response, described by the absolute gain $G$, differential gain $\gamma$, and differential phase $\phi$ of the receiver system, as a function of observing frequency.
The noise diode signal amplitude is further calibrated by observing it on and off a bright and unpolarized standard continuum calibrator radio source (flux calibrator).
Using flux calibrator observations during the calibration eliminates the assumption that the reference source illuminates both receptors equally.
For NANOGrav GBT observation with the GUPPI system, quasar B1442+101 was used as the continuum calibrator and observed approximately once per month.
One example IFA polarimetric response for calibrating the 800 MHz data is shown in panel (a) of Figure~\ref{fig:PRs}.
These instrumental responses for each pulsar observations are then used to calibrate the pulsar profiles.

It is important to note that this model relies on the assumption that the receptors are perfectly orthogonal and the noise diode signal is 100\% linearly polarized.
We have used the \texttt{pac} command in \texttt{PSRCHIVE}\footnote{\url{https://psrchive.sourceforge.net/manuals/}} \citep{Hotan2004} to perform IFA calibrations of the observed pulsar profiles.

\subsubsection{Measurement Equation Modeling}
\label{subsubsec:MEM}

Developed in \cite{vanStraten2004}, Measurement Equation Modeling (MEM) utilizes a long-track observation (or multiple observations) of a pulsar with strong linear polarization over a wide range of parallactic angles to generate a complete PR of the observing system at a fiducial epoch.
This method is based on the polarization measurement equation \citep{Hamaker2000}, which relates the measured Stokes parameters to the intrinsic pulsar polarization and is employed to determine the unknown instrumental response.

At first, a preliminary IFA polarization calibration is performed on the long-track observation to create a preliminary total integrated calibrated profile.
Thereafter, a specified set of pulse phase bins from this profile is chosen to be used as model constraints.
The Stokes parameters at these phase bins are fitted as a function of parallactic angles using the polarization measurement equation, and a best-fit solution for the complete instrumental PR at the fiducial epoch is calculated.
The complete PR is parameterized by the absolute gain $G$, differential gain $\gamma$, differential phase $\phi$, ellipticities of the two receptors $\epsilon_{k}$, and orientation of the receptor-1 with respect to receptor-0 $\theta_1$ as functions of observing frequency, when we use the \texttt{van04e18} parametrization \citep[equation 18 of][]{vanStraten2004}. \new{We have used this parametrization while calculating the MEM solutions for Rcvr\_800 (see panels (b, c, d) of Figure~\ref{fig:PRs}).
For the Rcvr1\_2 MEM solution, we have used the \texttt{bri00e19} model \citep{Britton2000}, where $\delta_{\theta, \chi}$ and $\sigma_{\chi}$ (defined by Equation~17 of \cite{Britton2000}) are used instead of $\epsilon_k$ and $\theta_1$ and the solution is shown in panel (e) of Figure~\ref{fig:PRs}.}
During this process, a flux calibrator observation is \new{optionally} used to constrain the mixing of Stokes I and V, and this breaks the degeneracy described in \cite{vanStraten2004}.

The \texttt{pcm} command in \texttt{PSRCHIVE} is used to generate the MEM PRs of the observing system from pulsar observations over a wide range of parallactic angles.
For calibrating Rcvr\_800 data, we used three long-track observations of PSR B1929+10 at 820 MHz on MJDs 56244, 56419, and 56608 to generate three separate MEM PR solutions at those three epochs.
This bright pulsar has well-known polarization characteristics and the data used here were acquired by \cite{Kramer2021} \new{to calibrate observations of} the double pulsar. 
A single long-track observation of PSR J1022+1001 on MJD 55671 is used to derive the MEM solution at 1500 MHz for calibrating the Rcvr1\_2 data.
Although PSR J1022+1001 has been found to exhibit long-term pulsar profile variation, we do not expect that to affect our result as the MEM solution is calculated from an observation of the pulsar on a single day \citep{Wahl2022}.

To perform the polarization calibration of a pulsar observation using the MEM method, the MEM solution from the fiducial epoch closest to the pulsar observation is first used to correct for the complete system response at that fiducial epoch.
Thereafter, the noise diode observation taken prior to the pulsar observation is employed to correct for any variations in differential gain or phase since the reference epoch.
The MEM polarization calibration method is based on the assumption that the receptor orientations and ellipticities, as well as the polarization of the reference source, do not vary significantly with time.

\subsubsection{Measurement Equation Template Matching}

Measurement Equation Template Matching (METM) was developed by \cite{vanStraten2013}, combining MTM with the MEM. 
In this method, the observation of a reference pulsar is matched with a well-calibrated template profile of that pulsar to obtain the best-fit METM model for the transformation between the template profile and the observation. 
The METM model can be used to fully represent the PR of the observing system at the observation epoch. 
This process can be repeated for multiple observations of the reference pulsar to obtain a complete PR of the observing system for each observation epoch. 
We can then use the METM models to calibrate other pulsar observations at those or nearby epochs. 
This method is valid under the assumption that the polarized emission from the reference pulsar remains constant over time.

In this paper, we use PSR B1937$+$21 as the reference pulsar for METM calibration due to its high brightness and well-known polarization characteristics.
Additionally, PSR~B1937$+$21 is observed in the same session as the three pulsars used in our analysis by NANOGrav, thus providing METM solutions for the exact epochs we require.
To generate METM solutions from the PSR~B1937+21 observations and perform polarization calibration on other pulsar observations, we followed the method outlined in \cite{Gentile2018}.
Initially, all B1937$+$21 observations are calibrated using the MEM method, following the procedure described in Section~\ref{subsubsec:MEM}. 
Among the calibrated profiles, two profiles are selected as template profiles---one for the 820 MHz band and another for the 1500 MHz band.
\new{We compared all our MEM-calibrated profiles of B1937$+$21 with the previously published polarization profiles \citep{DHM+2015}\footnote{\url{https://psrweb.jb.man.ac.uk/epndb/}} at the respective frequencies, selecting the profile that most closely matches the published one as the template.}
Thereafter, METM PRs are calculated for each observation epoch by comparing the MEM-calibrated profiles to the template profiles.
The \texttt{pcm} command from \texttt{PSRCHIVE} is used to obtain the METM PRs, and we used the option to normalize the Stokes parameters by the total invariant interval \citep{Britton2000} instead of calculating the absolute gain.
This decision was made because the reference pulsar B1937$+$21 exhibits pronounced scintillation \citep{Turner2024}, which could significantly bias the absolute gain due to variations in pulse intensity caused by scintillation across different frequencies.
\new{We further estimate the ionospheric RM contribution and subtract it from each METM PRs using the \texttt{pcmrm} command. This method of performing METM calibration has previously been used in \cite{Gentile2018} to generate accurate polarization profiles of 28 pulsars observed with the Arecibo telescope.}

Given that the observations used to obtain the METM PRs have already undergone calibration with the MEM-generated PR, it is appropriate to consider the METM-generated PRs as per-epoch corrections to the MEM-generated PR.
Accordingly, we apply these corrections to other pulsar observations that are already calibrated using MEM PRs. 
This is done by selecting the METM PR correction whose epoch is closest to the epoch of the observation that is to be calibrated.
\new{However, the differential gain and phase of the METM PR corrections are set to zero before being applied to the MEM-calibrated profiles. 
This is because the MEM-calibrated profiles have already been corrected for the receiver's differential gain and phase, which were determined using the noise diode observation conducted just prior to the pulsar observation.}
In our paper, this method of polarization calibration is denoted as MEM + METM, signifying the combination of both MEM and METM methods.
One example MEM+METM PR correction calculated from 800 MHz B1937+21 observation on MJD 56614 is shown in \new{panel (f)} of Figure~\ref{fig:PRs}.
\new{A brief summary of all the polarization calibration methods used in this paper is provided in Table~\ref{tab:pol_cal_methods}}.
In the following section, we discuss the process of generating TOAs from the calibrated profiles and performing timing analysis.

\begin{deluxetable*}{c c c}
\tablecaption{Summary of different polarization calibration methods used in the paper \label{tab:pol_cal_methods}}
\tablehead{
    \colhead{Method} &
    \colhead{Brief description} &
    \colhead{References}
    }
\startdata
IFA & \begin{minipage}[t]{0.5\linewidth} Corrects for the absolute gain ($G$), differential gain ($\gamma$), and differential phase ($\phi$) of the receiver system, calculated from a noise diode observation conducted just before the pulsar observation. \end{minipage} & \cite{StinebringCordes1984} \\
\hline
MEM & \begin{minipage}[t]{0.5\linewidth} Uses the full polarization response solutions, derived from long-track observations of a pulsar, to correct for $G$, $\gamma$, $\phi$, as well as the ellipticity ($\epsilon$) and non-orthogonality ($\theta$) of the feeds. \end{minipage} & \cite{vanStraten2004}\\
\hline
MEM + METM & \begin{minipage}[t]{0.5\linewidth} Applies METM polarization response correction in addition to MEM calibration to account for any potential variations in $\epsilon$ and $\theta$. \end{minipage} & \cite{vanStraten2013, Gentile2018}
\enddata
\end{deluxetable*}

\subsection{TOA Generation and Timing Analysis}
\label{subsec:toa_timing}

We started with the raw folded pulsar profiles and removed artifacts arising from the interleaved analog-to-digital converter scheme used by the GUPPI receiver system, as described in Section 2.3.1 of \cite{NG2021_12yr_timing}.
Thereafter, we performed the standard radio frequency interference (RFI) excision and calibrated the profiles using the three methods described above.
After calibrating the pulsar profiles, we performed additional steps of excising RFI from the calibrated profiles, following \cite{NG2023_15yr_timing}.
For MEM and MEM+METM calibrated profiles, we utilized \texttt{rmfit} to determine the optimal fit for the RM value associated with each profile and subsequently apply this calculated RM value to the profile header.
The clean calibrated profiles are then frequency-averaged into 64 channels (for both Rcvr\_800 and Rcvr1\_2 data) and time-averaged into subintegrations of up to 30 minutes.
Thereafter, narrowband template profiles are generated for each receiver band using the following steps.
Profiles associated with a specific pulsar and receiver band are aligned, weighted based on signal-to-noise ratio, and then summed to create a final averaged profile.
The resulting average profile is then `denoised' using wavelet decomposition and thresholding, and the whole process is iterated multiple times to converge on the final template \citep[for more details, see Section~3.1 of][]{Demorest2013}.
Finally, TOAs are obtained from folded pulse profile data by measuring the time shift of the observed profile relative to the template, following the standard approach used in pulsar timing analyses for decades \citep[e.g.,][]{Taylor1992}.
We have used only the total intensity profiles of the pulsars \new{and employed the \textit{Fourier domain with Markov chain Monte Carlo} (FDM) algorithm} while generating the TOAs.
The calculated TOAs are narrowband in nature, i.e, a separate TOA is measured for each frequency channel of the final profiles.
We used the \texttt{nanopipe}\footnote{\url{https://github.com/demorest/nanopipe}} \citep{Demorest2018} data processing pipeline, which in turn uses the \texttt{PSRCHIVE} pulsar data analysis software package, to perform these steps.
For more details on the TOA generation procedure, see Section~3 of \cite{NG2023_15yr_timing} and references therein.
It is crucial to emphasize that distinct templates have been generated for each polarization calibration method, and the corresponding template is employed when calculating the TOAs from profiles calibrated using different methods.

After obtaining the TOAs, we performed outlier removal on the files containing TOAs (\texttt{tim} files) for each polarization calibration method, following the procedures outlined in Section~3.3 of \cite{NG2023_15yr_timing}.
At first, an initial timing solution (timing model parameter file or \texttt{par} file) was derived using the \texttt{tim} file, which contains both Rcvr\_800 and Rcvr1\_2 TOAs. Subsequently, the initial \texttt{par} and \texttt{tim} files were passed to an automated outlier analysis pipeline, which removed TOAs with outlier probabilities exceeding $0.1$ from the \texttt{tim} file.
Furthermore, TOAs from a specific folded profile were excluded if a substantial percentage of them were flagged as outliers.
In addition, any TOA associated with profile data that did not meet the signal-to-noise ratio threshold (S/N $>8$) was removed from the \texttt{tim} \new{file} (see \citep{NG2023_15yr_timing} for more details).
The resulting excised \texttt{tim} file was then used for the subsequent timing analysis.

We employed the PINT \citep{PINT2019_software} pulsar timing software \citep{Luo2021_PINT} and adopted the procedure outlined in the NANOGrav 15-year data release \citep{NG2023_15yr_timing} for our timing analysis.
We started with the NANOGrav 15-year \texttt{.par} files and the human-readable configuration (\texttt{.yaml}) files, and used the standardized Jupyter notebooks to automate our timing procedure.
The timing analysis was conducted using the JPL DE440 solar system ephemeris \citep{Park2021} and the TT(BIPM2021) timescale.
Pulsar timing involves comparing observed TOAs with TOAs predicted by a timing model, yielding timing residuals. 
\new{The timing model encompasses various parameters (e.g., pulsar period and period derivative, dispersion measure, pulsar sky location, orbital parameters if the pulsar is in a binary system) representing different physical effects influencing pulse arrival times \citep{LorimerKramer2004hpa, Hobbs+2006}.}
During the pulsar timing, best fit values of the timing model parameters are calculated to minimize the RMS of the timing \new{residuals}.
Here we briefly describe the different timing parameters used in our analysis.

For each pulsar, we fit for two spin parameters (rotational frequency and frequency derivative) and five astrometry parameters (two-dimensional sky position and proper motion, and parallax).
For PSRs J1643$-$1224 and J1909$-$3744, we also fit binary parameters describing \new{an orbit} with a companion star.
For PSR J1643$-$1224, we employ the DD binary model \citep{DD1985}, incorporating the following six parameters: orbital period $P_b$, projected semi-major axis $x$ and its time derivative $\Dot{x}$, orbital eccentricity $e$, longitude of periastron $\omega$, and epoch of periastron passage $T_0$.
The ELL1 binary model \citep{Lange_ELL1_2001} is used for PSR J1909$-$3744, where in addition to $P_b$, $x$, and $\Dot{x}$, we have incorporated $\Dot{P}_b$, the companion mass $m_2$, orbital inclination parameter $\sin{i}$, two Laplace–Lagrange parameters ($\epsilon_1$, $\epsilon_2$) and the epoch of the ascending node $T_{asc}$.

The variation in the interstellar dispersion measure (DM) is mitigated using the DMX model, a piecewise constant function, with each DMX parameter describing the offset from a nominal fixed value.
We also fit for additional time-independent but frequency-dependent delays on a per-pulsar basis using ``FD" parameters (see \cite{NG2015_9yr_timing}).
The FD parameters account for time offsets resulting from disparities between the observed pulse shape at a specific frequency and the template shape used in timing, and the number of these parameters included is determined via the F-test procedure discussed in Section~4.1.2 of \cite{NG2023_15yr_timing}.
Furthermore, we incorporate ``JUMP"s to address unknown phase offsets between data observed by different receivers.

The scripts used for performing the polarization calibration and TOA generation, along with all the MEM and METM PR solutions can be found in \cite{psrcal_script2024}.
The \texttt{PINT}-based timing and outlier analysis packages, along with relevant example Jupyter notebooks used in our analysis, are available in \texttt{PINT\_pal} \citep{pint_pal2023}.

\section{Results} 
\label{sec:results}

\begin{figure*}
    \centering
    \includegraphics[width=0.45\textwidth]{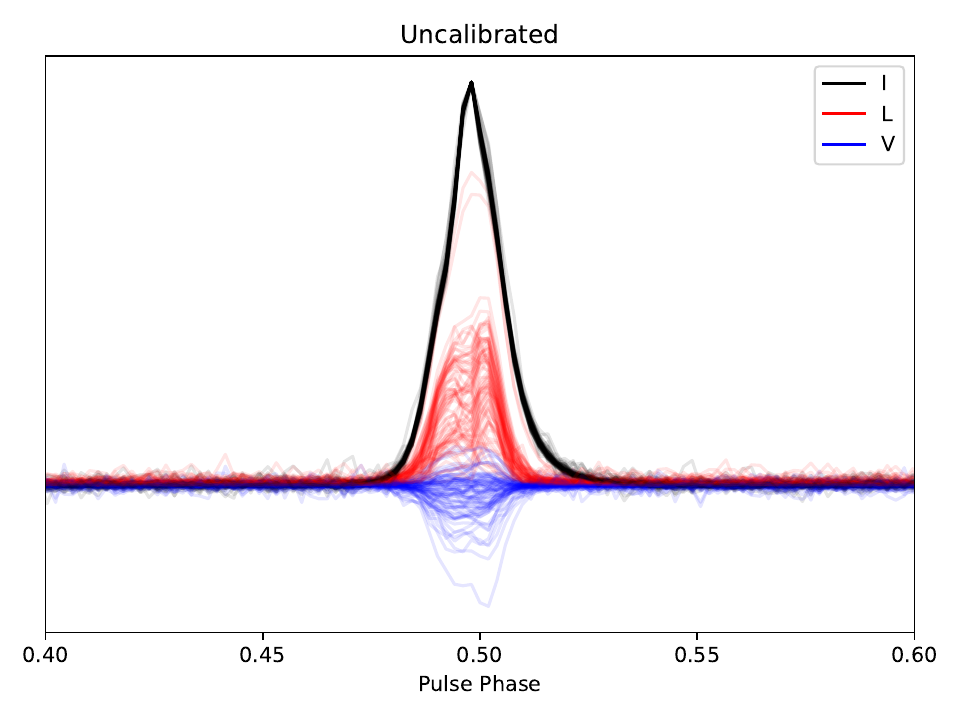}
    \includegraphics[width=0.45\textwidth]{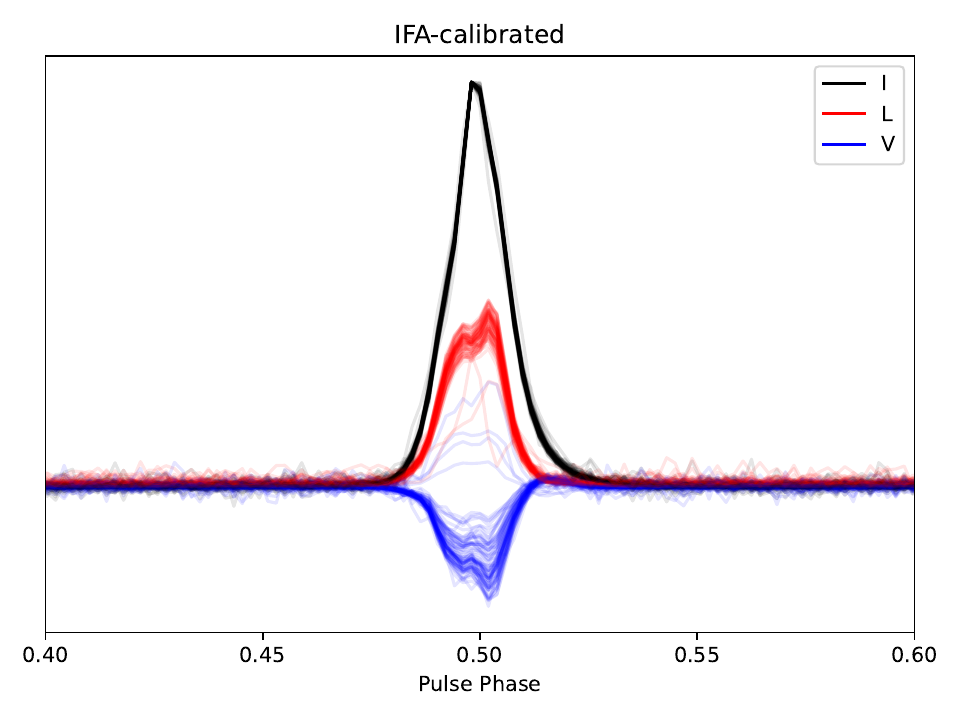}\\
    \includegraphics[width=0.45\textwidth]{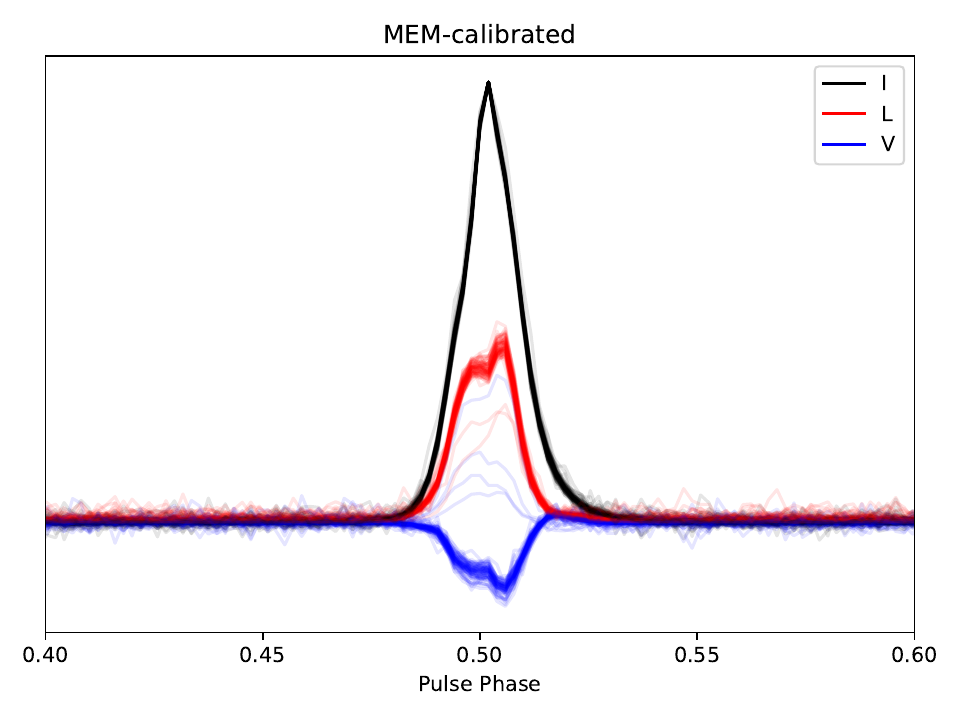}
    \includegraphics[width=0.45\textwidth]{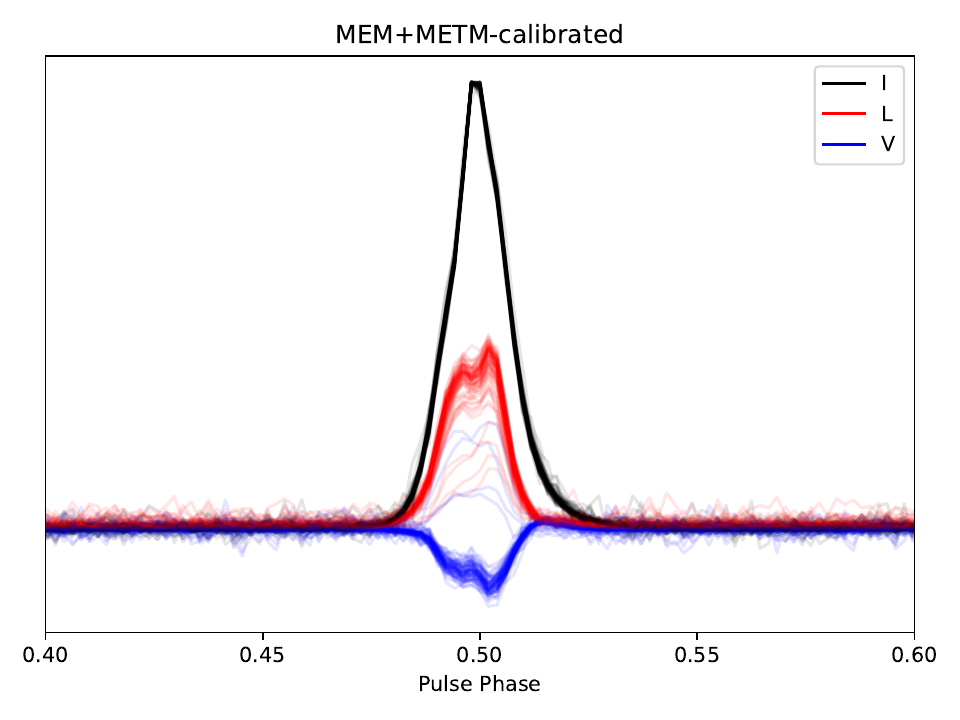}
    \caption{Uncalibrated and calibrated (using different methods) polarization profiles of PSR J1909$-$3744 in the Rcvr1\_2 band (1500 MHz).
    The calibration method used to obtain the profiles are denoted in the plot titles.
    In each panel, the black, red, and blue lines indicate the total intensity (I), linear polarization (L), and circular polarization (V), respectively.
    Using the IAU’s circular polarization sign convention, right-handed circular polarization is positive and left-handed circular polarization is negative.}
    \label{fig:J1909_1500_profs}
\end{figure*}

In this section, we present the outcomes of our diverse polarization calibration processes and the corresponding timing analysis for PSRs J1643$-$1224, J1744$-$1134, and J1909$-$3744. 
To illustrate the varied results from different polarization calibration procedures, Figure~\ref{fig:J1909_1500_profs} displays all the Rcvr1\_2 profiles of PSR J1909$-$3744 from different observation epochs.
Each panel in the figure represents profiles obtained through distinct polarization calibration methods, with the black, red, and blue lines indicating the total intensity (I), linear polarization (L), and circular polarization (V), respectively. 
Additionally, we include the original uncalibrated profiles (after RFI excision) in the top-left panel of the figure for comparison.
We observe from Figure~\ref{fig:J1909_1500_profs} that, in the case of uncalibrated profiles, although the total intensity profiles closely align across different epochs, significant variability exists in both the linear and circular polarization profiles from epoch to epoch.
Polarization calibration proves effective in mitigating variations in linear polarization profiles, with only a very few epochs showing exceptions, across all three calibration methods. 
The same holds true for circular polarization, although the MEM and MEM+METM methods seem to do a better job compared to the IFA polarization calibration.
Similar trends are seen for the Rcvr\_800 observing frequency and for other pulsars, in general.
However, the calibration processes are not always able to reasonably mitigate the epoch-to-epoch variations in the linear and circular polarization profiles.
This is particularly prominent for PSR J1744$-$1134 where we see variations (at different levels) in the circular polarization even after performing the polarization calibration.
The uncalibrated and calibrated profiles for J1909$-$3744 Rcvr\_800 observations and of PSRs J1643$-$1224 and J1744$-$1134 are shown in the Appendix~\ref{app:profiles}.

\begin{deluxetable*}{c c c c c c c c}
\tablecaption{Timing analysis statistics for different polarization calibration methods \label{tab:timing_stats}}
\tablehead{
    \colhead{Pulsar} &
    \colhead{Method} &
    \colhead{\phantom{X}N$_{\text{TOA}}$\phantom{X}} &
    \colhead{$\sigma_{\text{med}}$} &
    \colhead{Median} &
    \colhead{\phantom{X}RMS ($\mu$s)\phantom{X}} &
    \colhead{\phantom{X}WRMS ($\mu$s)\phantom{X}} &
    \colhead{Reduced}\\
    &
    &
    &
    ($\mu$s) &
    S/N &
    \colhead{\footnotesize All TOAs/Epoch-avg.} &
    \colhead{\footnotesize All TOAs/Epoch-avg.} &
    \colhead{chi-squared}
    }
\startdata
\multirow{4}{*}{J1643$-$1224} & Uncalibrated & 7352 & 2.02 & 111.62 & 4.126 / 0.763 & 3.464 / 0.780 & 3.664 \\
& IFA & 7172 & 1.99 & 114.10 & 3.033 / 0.822 & 2.332 / 0.782 & 1.682 \\
& MEM & 7019 & 2.02 & 111.85 & 3.686 / 0.830 & 2.909 / 0.809 & 2.520 \\
& MEM+METM & 6850 & 2.14 & 104.52 & 3.849 / 0.859 & 2.966 / 0.816 & 2.233 \\
\hline
\multirow{4}{*}{J1744$-$1134} & Uncalibrated & 7467 & 0.997 & 76.93 & 2.823 / 0.749 & 0.655 / 0.277 & 2.892 \\
& IFA & 7433 & 0.996 & 77.41 & 2.786 / 0.544 & 0.454 / 0.169 & 1.475 \\
& MEM & 7190 & 1.07 & 72.08 & 2.959 / 0.470 & 0.572 / 0.179 & 2.030 \\
& MEM+METM & 6715 & 1.12 & 68.54 & 2.945 / 0.575 & 0.654 / 0.201 & 2.319 \\
\hline
\multirow{4}{*}{J1909$-$3744} & Uncalibrated & 9643 & 0.498 & 55.06 & 1.187 / 0.139 & 0.176 / 0.051 & 1.464 \\
& IFA & 9488 & 0.502 & 54.54 & 1.193 / 0.089 &  0.151 / 0.034 & 1.108 \\
& MEM & 9163 & 0.513 & 53.10 & 1.208 / 0.101 & 0.164 / 0.035 & 1.229 \\
& MEM+METM & 8760 & 0.543 & 50.07 & 1.223 / 0.108 & 0.180 / 0.036 & 1.294 \\
\enddata
\end{deluxetable*}

In Table~\ref{tab:timing_stats}, various quantities and statistics related to the timing analysis for each pulsar are presented. 
The columns, from left to right, display the pulsar name, polarization calibration method, number of TOAs (N$_{\text{TOA}}$), median uncertainty of the TOAs ($\sigma_{\text{med}}$), median signal-to-noise ratio (S/N) of each sub-band of the pulsar profiles from which the TOAs are calculated, RMS (all TOAs/epoch averaged) and weighted RMS (all TOAs/epoch averaged) of the timing residuals, and the reduced chi-squared of the timing solution, respectively.
For each pulsar, the first row shows the results when we do not perform any polarization calibration before generating the TOAs, while the subsequent three rows represent results for IFA, MEM, and MEM+METM calibration.

For all three pulsars, we observe that the number of TOAs utilized in the timing analysis (post-outlier analysis) is highest when no polarization calibration is performed. 
This is primarily because polarization calibration can occasionally corrupt a few frequency channels, typically due to presence of \new{residual RFI} in the reference noise diode or the reference pulsar observation.
These corrupted channels are subsequently either zapped during RFI excision or the corresponding TOAs are excised during the outlier analysis.
Additionally, the calibration process can fail for a few channels due to the absence of a solution for those channels in the polarization response calculated using the MEM or MEM+METM methods.
Further, the MEM+METM calibration process failed for five J1744$-$1134 Rcvr\_800 observations due to a mismatch in the center frequency and observation bandwidth between the observed profiles and METM PRs.
We also see from the Table~\ref{tab:timing_stats} that the calculated median TOA uncertainties for the \texttt{.tim} files generated from the uncalibrated and IFA-calibrated profiles are very similar and have the lowest values.
Conversely, the $\sigma_{\text{med}}$ values are highest for the MEM+METM calibration, while the MEM calibration falls in the middle in this regard.
This trend aligns with the observed variations in the S/N between the uncalibrated and different calibrated profiles.
Typically, the S/N is highest for uncalibrated and IFA-calibrated profiles, lower for MEM-calibrated profiles, and lowest for MEM+METM-calibrated profiles.
\new{We note that this is in contradiction with similar studies for other telescopes, e.g., \cite{Guillemot2023} found MEM calibration led to higher S/Ns compared to IFA calibration for pulsar data taken with the Nan{\c{c}}ay Radio Telescope.}

We now shift our focus to examining various statistics of the timing solutions obtained for different calibration methods applied to each pulsar.
Upon inspecting the reduced chi-squared values, it becomes evident that the TOAs generated from all three calibration methods exhibit a superior fit to the timing model compared to the uncalibrated TOAs.
Among the different calibration methods, the IFA calibration yields reduced chi-squared values closest to unity for all three pulsars. 
However, there are variations observed while comparing the MEM and MEM+METM calibration methods.
For PSR J1643$-$1224, the reduced $\chi^2$ value is closer to unity for the MEM+METM calibration method compared to the MEM method, whereas for PSRs J1744$-$1134 and J1909$-$3744, the opposite holds true.

In Table~\ref{tab:timing_stats}, we also present two sets of values for the RMS and Weighted RMS (WRMS) of the timing solutions: one labeled as \textit{All TOAs}, and the other as \textit{Epoch-averaged}.
As their names suggest, the \textit{All TOAs} RMS and WRMS values are computed using all the sub-banded TOAs. 
Conversely, the \textit{Epoch-avg.} values are derived by averaging the TOAs from all sub-bands for a specific epoch to generate a single TOA.
Upon comparing the RMS and WRMS values across different calibration methods, we observe variations in the trend among different pulsars.
However, in most cases, the IFA calibration method consistently provides the lowest RMS and WRMS values compared to other methods.
Therefore, it is evident from Table~\ref{tab:timing_stats} that overall, the IFA polarization calibration works best for the GBT data taken with the GUPPI receiver system in terms of timing analysis. 
The other calibration methods, such as MEM and MEM+METM, offer improvements over performing no calibration, yet they demonstrate inferior performance compared to the IFA calibration method for our dataset.

\section{Discussions} 
\label{sec:discussion}

When a pulsar is observed with a radio telescope, instrumental artifacts can distort its total intensity profile, resulting in significant systematic timing errors. 
Therefore, ensuring accurate polarization calibration of the observed pulsar profiles is crucial for achieving high-precision timing, which is fundamental for PTA experiments. 
In this paper, we compare the performance of three different polarization calibration methods---IFA, MEM, and MEM+METM---using GBT observations of PSRs J1643$-$1224, J1744$-$1134, and J1909$-$3744 with the GUPPI receiver system.
Our findings indicate that all three calibration methods improve timing precision compared to scenario where no polarization calibration is conducted. 
This improvement is expected as polarization calibration corrects for instrumental response, resulting in more stable intrinsic total intensity pulse profiles. 
Consequently, this leads to more accurate and precise estimation of TOAs, thereby enhancing timing precision.

Based on previous studies \citep{vanStraten2013, ManchesterHobbs+2013, Rogers2020}, we anticipated that the MEM and MEM+METM calibration methods would yield better timing performance compared to the IFA calibration. 
This expectation also arises from the fact that, unlike MEM and MEM+METM methods, the IFA calibration does not correct for the ellipticities and non-orthogonality of the receiver feeds.
Contrary to our expectations, however, it was found that the IFA-calibrated data produced TOAs with the smallest errors.
In order to understand these results, it is important to recall that each of the polarization calibration methods operates based on distinct assumptions regarding the receiver and reference noise diode systems. 
For IFA, the assumptions are that the receptors are perfectly orthogonally polarized and the noise diode is 100\% linearly polarized.
The MEM calibration method assumes that the orientations and ellipticities of the receptors, as well as the polarization of the reference noise diode, do not significantly vary over time. 
Lastly, the METM calibration requires the polarization profiles of the reference pulsar to be stable and not subject to variation over time.

Upon examining the MEM PR solutions depicted in Figure~\ref{fig:PRs}, we observe that the ellipticities and non-orthogonality of the receptors (represented by $\epsilon_k$ and $\theta_1$) are \new{non-negligible and vary with frequency. 
Moreover, the values of $\epsilon_k$ and $\theta_1$, as well as their frequency-dependent trends, differ from epoch to epoch, as evident from the three MEM solutions at 800 MHz (see Figure~\ref{fig:PRs}).
Therefore, the assumption in the IFA approach of perfectly orthogonally polarized receptors does not seem to be entirely accurate for the receiver systems at the GBT, and is likely to degrade the accuracy of the polarization calibration to some degree.}
Determining whether the reference noise diode is inherently 100\% linearly polarized presents a challenge as the instrumental response must be decoupled from the noise diode observations.
PRs computed in the IFA approach assume the noise diode to be 100\% linearly polarized, rendering them unsuitable for correcting the instrumental response in the noise diode observations.
\new{However, during the generation of MEM PR solutions, the intrinsic Stokes parameters of the noise diode signal are modeled alongside the receptor parameters as part of the MEM fitting process, allowing for the determination of the intrinsic polarization properties of the noise diode.}

\begin{figure}
    \centering
    \includegraphics[trim={30, 10, 70, 50}, clip, width=0.75\textwidth]{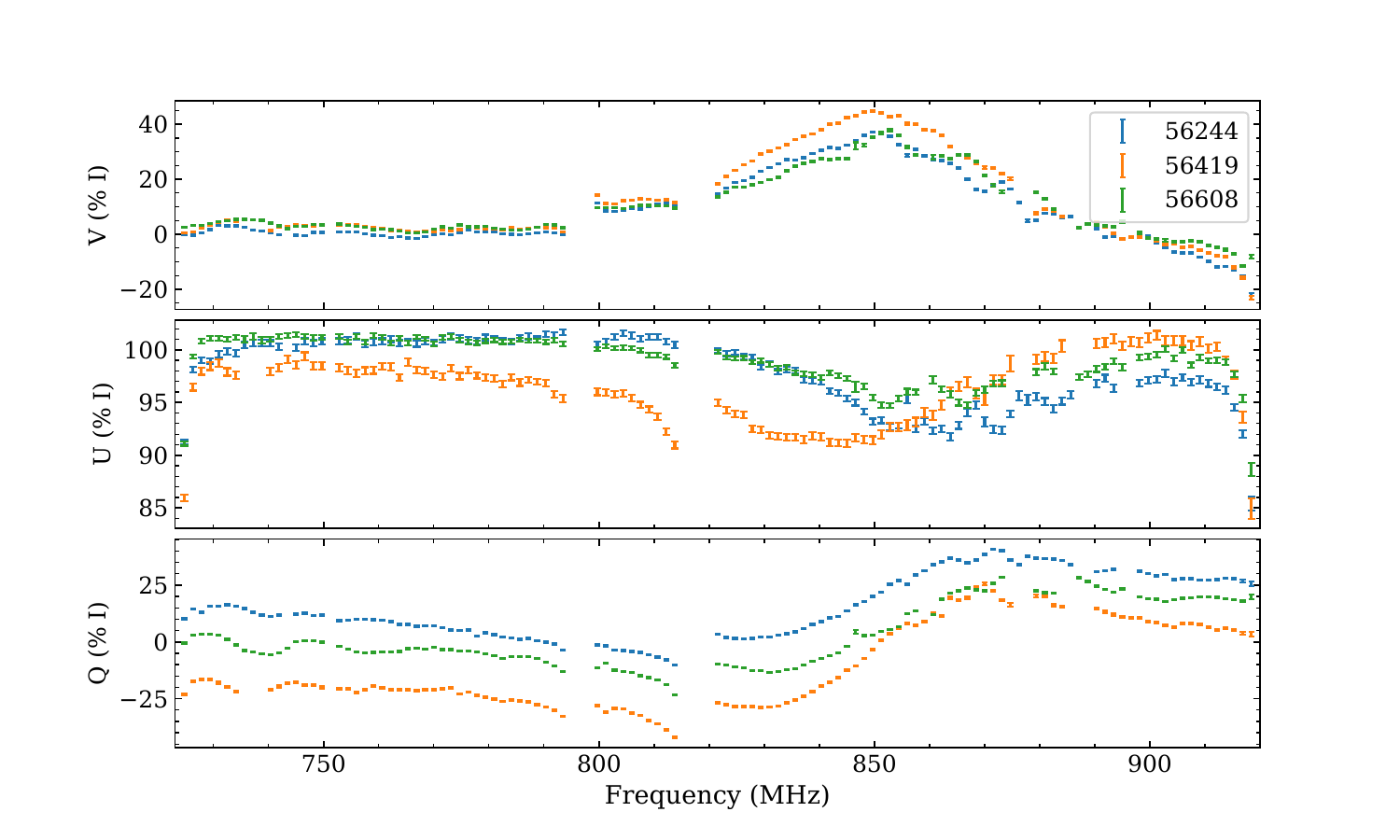}
    \caption{Intrinsic Stokes parameters of the noise diode reference signal for Rcvr\_800, plotted as a function of observing frequency for three distinct epochs: MJDs 56244, 56419, and 56608.
     The modeled values of Stokes $Q$, $U$, and $V$ are expressed as percentages of the total intensity of the reference source.}
    \label{fig:cal_stokes}
\end{figure}

In Figure~\ref{fig:cal_stokes}, we show the intrinsic Stokes parameters of the noise diode signal in the Rcvr\_800 band by presenting the fractional Stokes $Q$, $U$, and $V$, modeled \new{along with} the MEM PR solutions on MJDs 56244, 56419, and 56608.
For an ideal reference source illuminating both receptors equally, $U$ should register at 100\%, while $Q$ and $V$ should remain at zero across the entire frequency range.
However, since we incorporate a flux calibrator observation during our IFA calibration, the assumption of equal illumination on both receptors is eliminated, and therefore, the noise diode signal only needs to be 100\% linearly polarized.
Inspection of Figure~\ref{fig:cal_stokes} reveals that across all three epochs, the reference signal is $\sim 95-100$\% linearly polarized for the majority of the band. 
It is also evident from the figure that the intrinsic polarization characteristics of the reference noise diode exhibit variability from epoch to epoch.
At the Rcvr1\_2 observing frequencies, we observe from the single MEM solution available on MJD 55671 that the reference noise diode signal consists of $\sim 90$\% linear polarization.
However, due to the limited availability of MEM solutions for only one epoch, we lack information regarding the temporal variations in the polarization of the reference noise diode for the Rcvr1\_2 system.

Therefore, we see that for both the IFA and MEM approaches for polarization calibration, the assumptions made regarding the feeds and the reference noise diode signal are not entirely applicable to the GBT receivers.
\new{Additionally, we used only three and one MEM PR solutions for calibrating four years of Rcvr\_800 and Rcvr1\_2 data, respectively.
This would have been effective only if the system response, specifically the ellipticity and non-orthogonality of the feeds, remained stable, and the polarization properties of the noise diode were consistent over that time.
However, as seen in Figure~\ref{fig:PRs}, the three MEM PR solutions for Rcvr\_800 show variability in ellipticity and non-orthogonality of the feed over time, which likely impacted the accuracy of our MEM calibration.
The IFA calibration should also be affected by variability in the ellipticity and non-orthogonality of the receiver feeds, as we assume these parameters are zero in this method.
Interestingly, our results indicate that the TOAs derived from IFA-calibrated pulse profiles exhibit fewer systematic errors compared to those from MEM calibration.
This leads us to believe that, while the non-zero ellipticities and non-orthogonality of the receiver feeds, along with the reference signal being $\gtrsim 90$\% linearly polarized, impact the timing performance of the IFA-calibrated data, the temporal variations in the polarization of the noise diode signal more significantly degrade the timing performance of MEM-calibrated data.}
However, further in-depth investigations are necessary to validate these conclusions and quantify how the deviations from the underlying assumptions for different polarization calibration methods affect the accuracy of the calibration.

For MEM+METM polarization calibration, we have used the PSR B1937$+$21 as the reference pulsar to generate the per-epoch corrections to the MEM-generated PRs.
However, we observed that the MEM+METM calibration typically performs worse than both the IFA and MEM calibrations.
We suspect that this occurs because the polarization profiles of PSR B1937$+$21 do not remain stable over time.
One potential cause of this instability could be variable scatter broadening, leading to temporal profile variations, particularly at lower frequencies \citep[see][]{Brook2018}.

In previous NANOGrav data releases \citep[e.g.,][]{NG2015_9yr_timing, NG2021_12yr_timing, NG2023_15yr_timing}, the IFA approach has been employed to calibrate observed pulsar profiles.
Our analysis of three pulsars with data obtained with GBT Rcvr\_800 and Rcvr1\_2 receivers, coupled with the GUPPI backend system, indicates that IFA polarization calibration yields the best results.
Therefore, it is recommended to continue using IFA polarization calibration for future NANOGrav data sets until any potential changes in the GBT receiver systems.
However, conducting similar analyses for additional pulsars and also using data obtained from other telescopes, such as Arecibo and the Very Large Array (VLA), and with other receiver systems would be valuable to validate these findings.
\new{Furthermore, mitigating the scatter broadening in the observed profiles of PSR B1937+21 prior to generating the METM PR corrections could prove beneficial and will be explored in future studies. Current efforts to develop a cyclic spectroscopic backend system for GBT pulsar observations also aim to facilitate this by calculating the impulse response function, which will be used to mitigate scatter broadening effects in pulsar profiles in the near future \citep{TurnerStinebring+2023}.}
Additionally, we plan to investigate whether using matrix template matching instead of the standard total intensity to generate TOAs from calibrated profiles alters our results.

\section*{Author Contributions}
L.D. carried out the analyses and prepared the text, figures, and tables.
M.A.M. and H.M.W. helped with the development of the framework.
P.B.D., S.B.S., W.F., J.G., and R.J.J. helped with useful inputs and discussions.
The rest of the authors (including M.A.M., H.M.W., and P.B.D.) contributed to the collection and analysis of the NANOGrav 12.5 yr data set.

\section*{Acknowledgments}
\new{We thank the anonymous referee for their valuable comments and suggestions, which have helped us to improve the paper.}
We also thank Willem van Straten for useful comments and discussions.
This work has been carried out as part of the NANOGrav collaboration, which receives support from the National Science Foundation (NSF) Physics Frontiers Center award numbers 1430284 and 2020265.
The Green Bank Observatory is a facility of the NSF operated under a cooperative agreement by Associated Universities, Inc. 
L.D. is supported by a West Virginia University postdoctoral fellowship.
P.R.B. is supported by the Science and Technology Facilities Council, grant number ST/W000946/1.
H.T.C. acknowledges support from the U.S. Naval Research Laboratory, where basic research in pulsar astronomy is supported by ONR 6.1 funding.
T.D. and M.T.L. are supported by an NSF Astronomy and Astrophysics Grant (AAG) award number 2009468.
E.C.F. is supported by NASA under award number 80GSFC21M0002.
D.R.L. and M.A.M. are supported by NSF \#1458952.
M.A.M. is supported by NSF \#2009425.
The Dunlap Institute is funded by an endowment established by the David Dunlap family and the University of Toronto.
T.T.P. acknowledges support from the Extragalactic Astrophysics Research Group at E\"{o}tv\"{o}s Lor\'{a}nd University, funded by the E\"{o}tv\"{o}s Lor\'{a}nd Research Network (ELKH), which was used during the development of this research.
N.S.P. was supported by the Vanderbilt Initiative in Data Intensive Astrophysics (VIDA) Fellowship.
S.M.R. and I.H.S. are CIFAR Fellows.
Pulsar research at UBC is supported by an NSERC Discovery Grant and by CIFAR.


%

\vspace{5mm}
\facilities{Green Bank Telescope}


\software{\texttt{PINT} \citep{PINT2019_software}, \texttt{PSRCHIVE} \citep{Hotan2004}, \texttt{ENTERPRISE} \citep{EllisVallisneri+2017}, \texttt{numpy} \citep{HarrisMillman+2020}, \texttt{matplotlib} \citep{Hunter2007}.}

\appendix

\section{Calibrated and uncalibrated profiles}
\label{app:profiles}

\begin{figure*}
    \centering
    \includegraphics[width=0.45\textwidth]{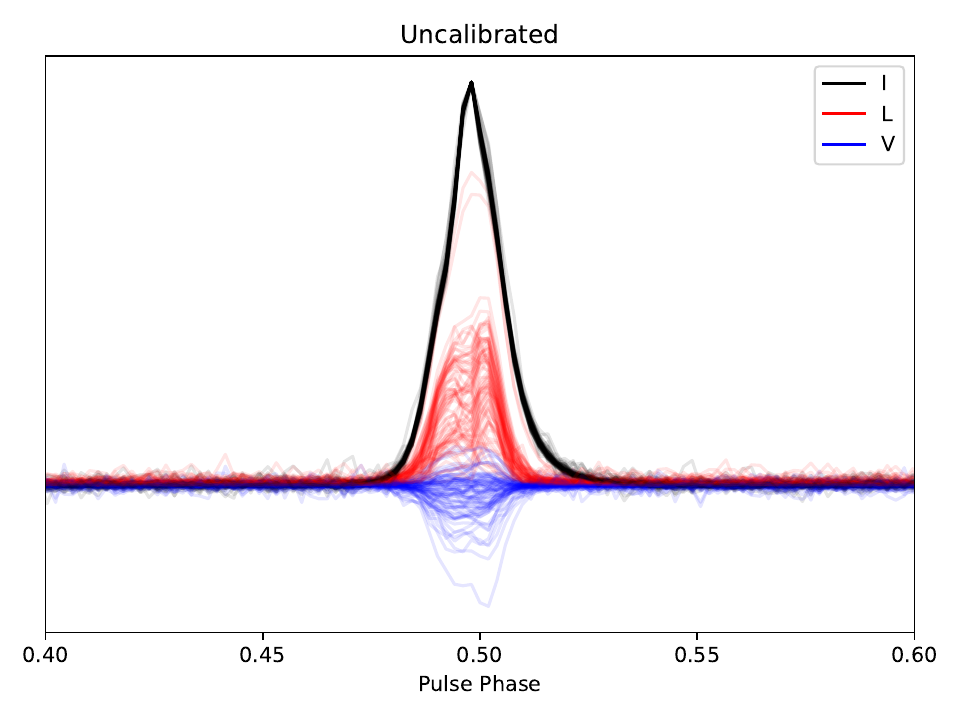}
    \includegraphics[width=0.45\textwidth]{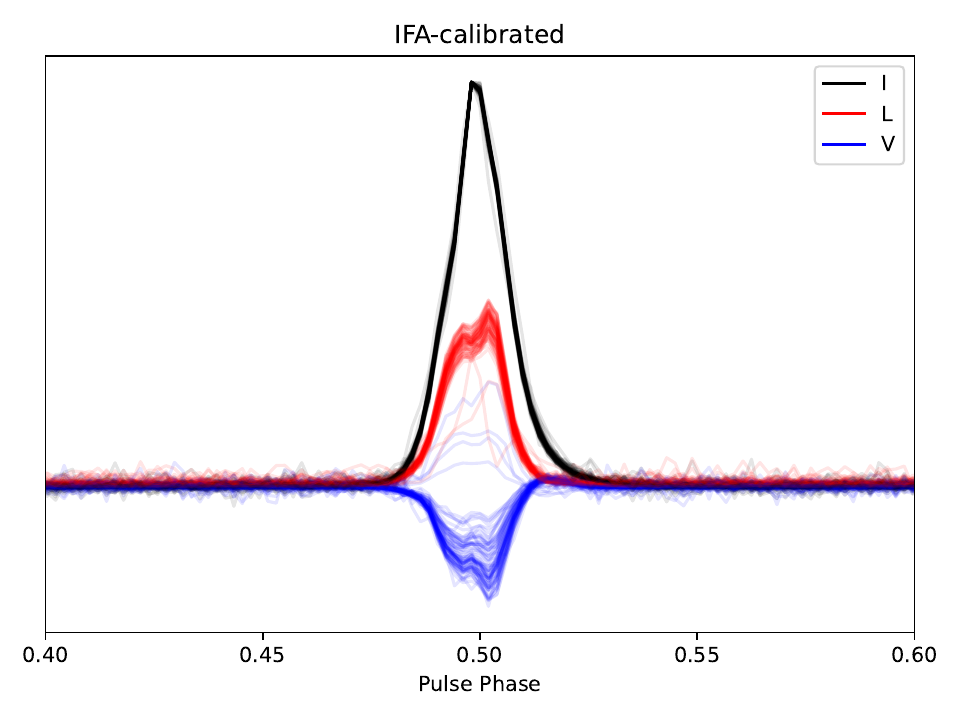}\\
    \includegraphics[width=0.45\textwidth]{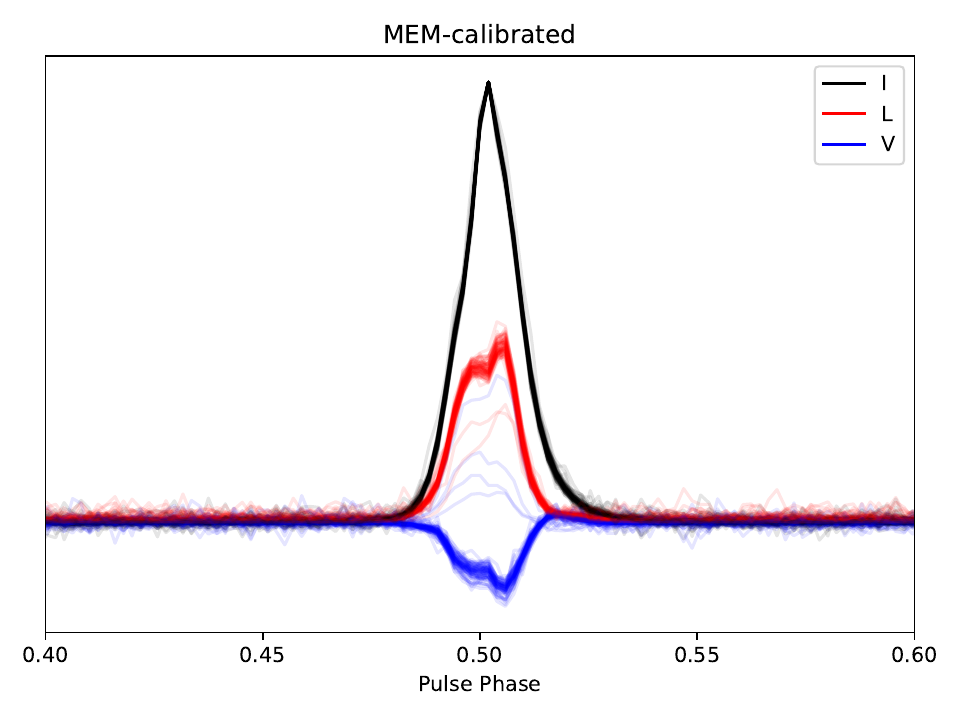}
    \includegraphics[width=0.45\textwidth]{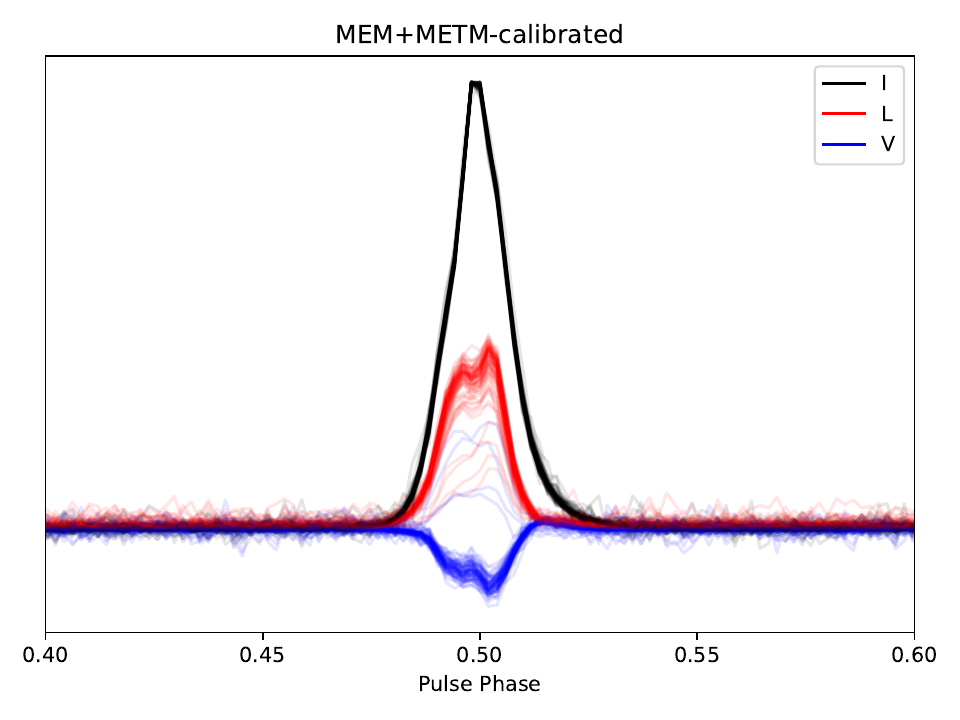}
    \caption{Polarization profiles for PSR J1909$-$3744 observed with the GUPPI 800 MHz receiver system at GBT. Both uncalibrated profiles and profiles obtained by different calibration methods are shown.}
    \label{fig:J1909_800_profs}
\end{figure*}

\begin{figure*}
    \centering
    \includegraphics[width=0.4\textwidth]{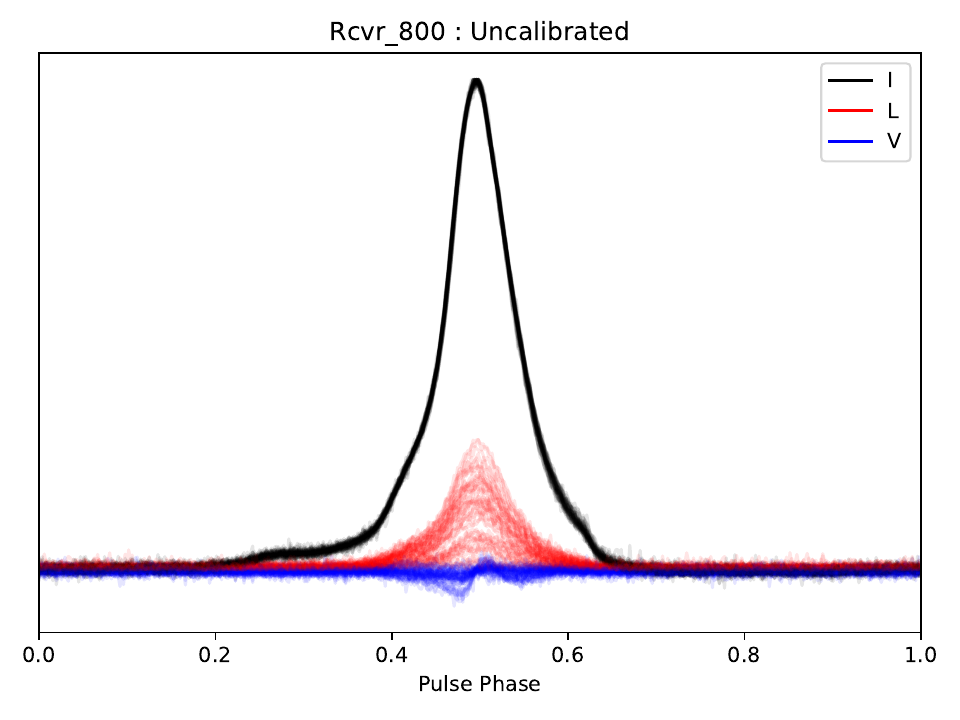}
    \includegraphics[width=0.4\textwidth]{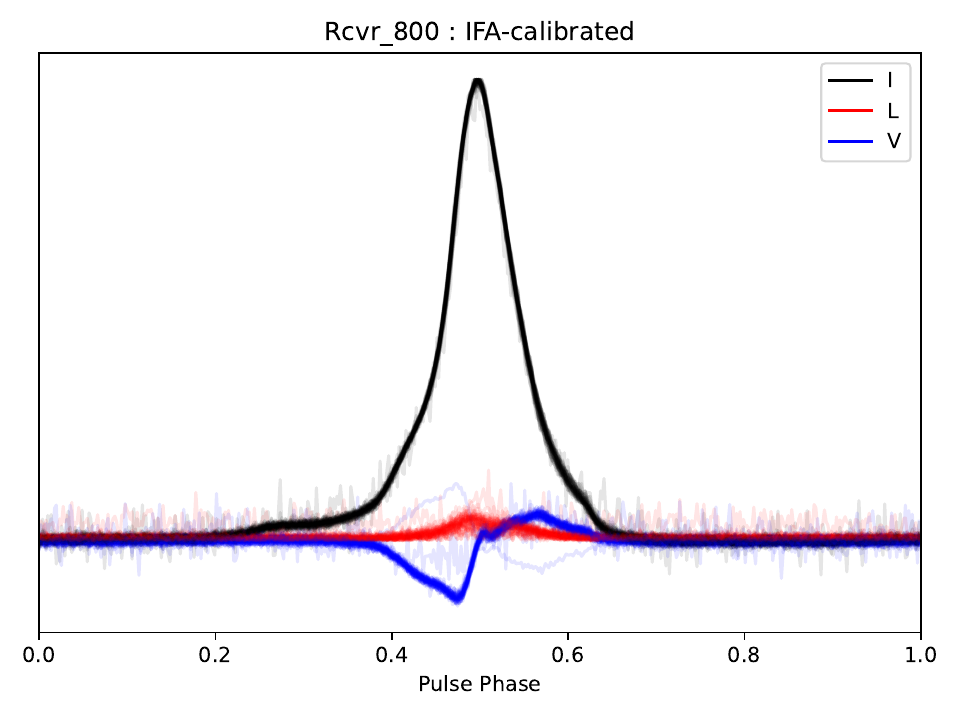}\\
    \includegraphics[width=0.4\textwidth]{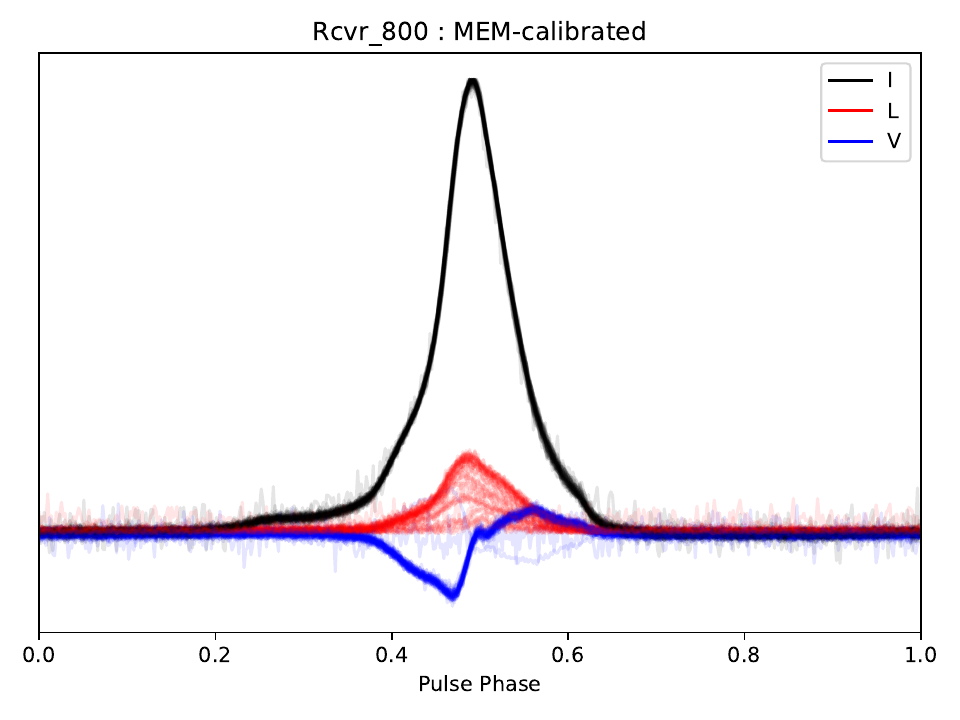}
    \includegraphics[width=0.4\textwidth]{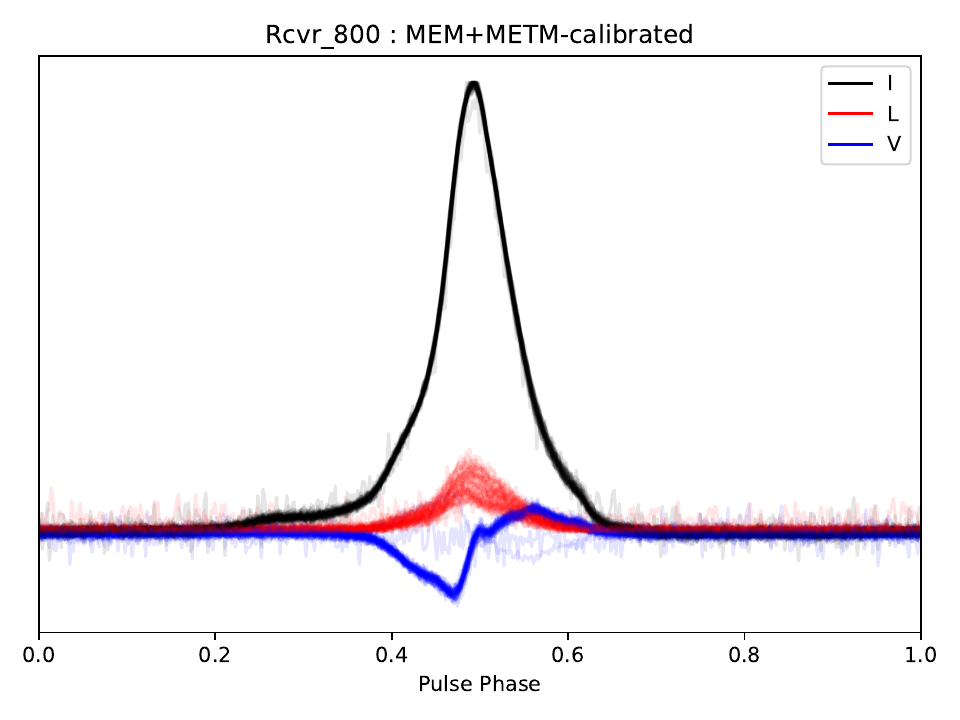}\\
    \includegraphics[width=0.4\textwidth]{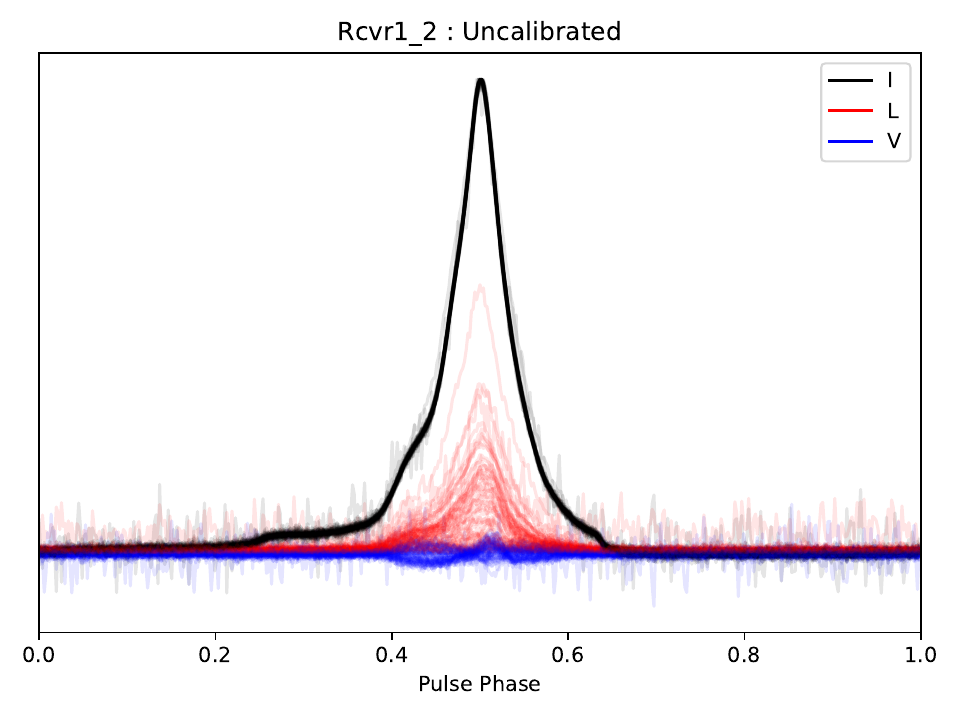}
    \includegraphics[width=0.4\textwidth]{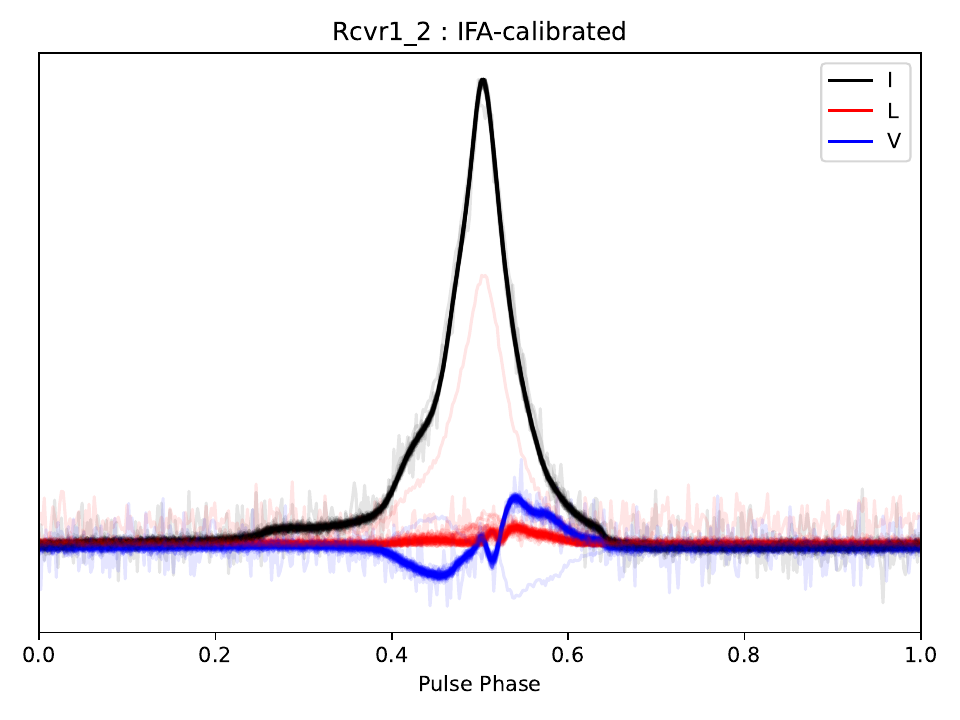}\\
    \includegraphics[width=0.4\textwidth]{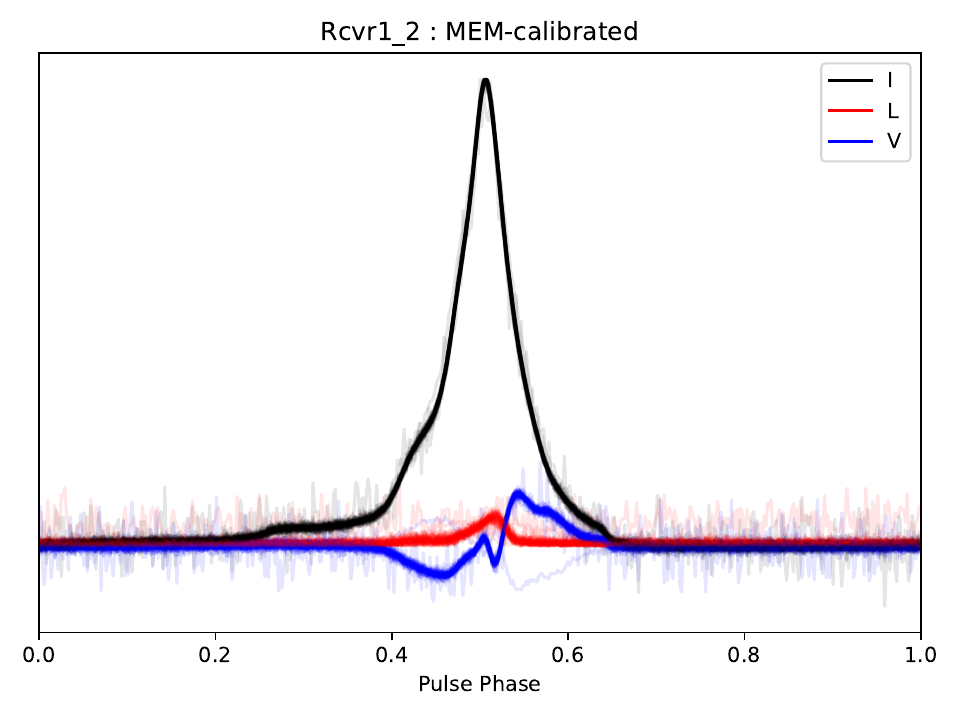}
    \includegraphics[width=0.4\textwidth]{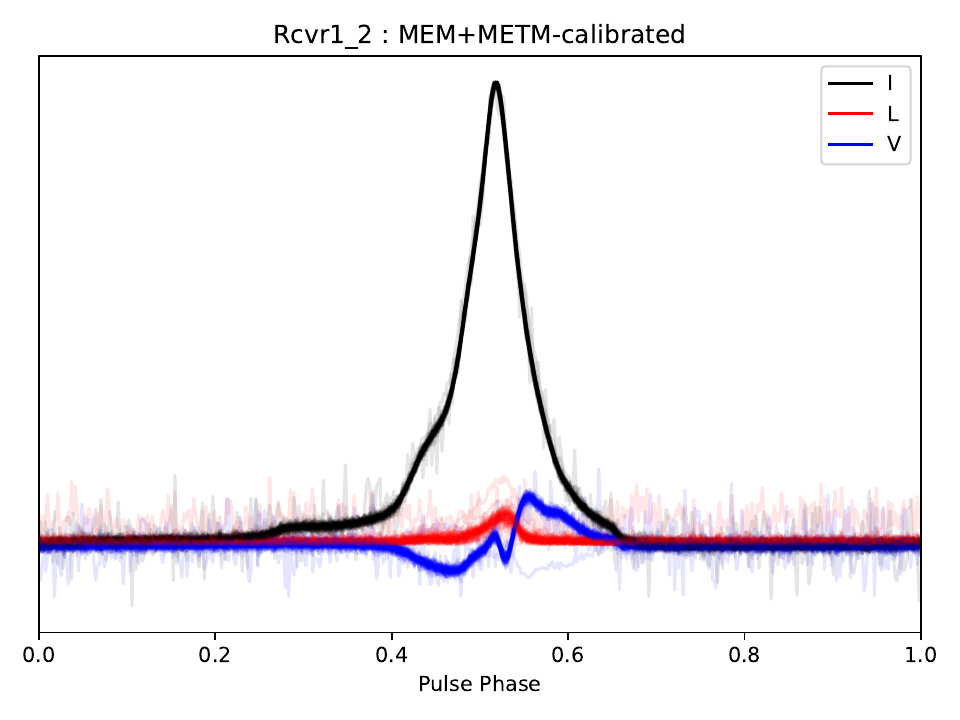}\\
    \caption{Uncalibrated and different calibrated profiles for J1643$-$1224 observed with Rcvr\_800 (800 MHz) and Rcvr1\_2 (1500 MHz) receivers and GUPPI backend system at the GBT.}
    \label{fig:J1643_profs}
\end{figure*}

\begin{figure*}
    \centering
    \includegraphics[width=0.4\textwidth]{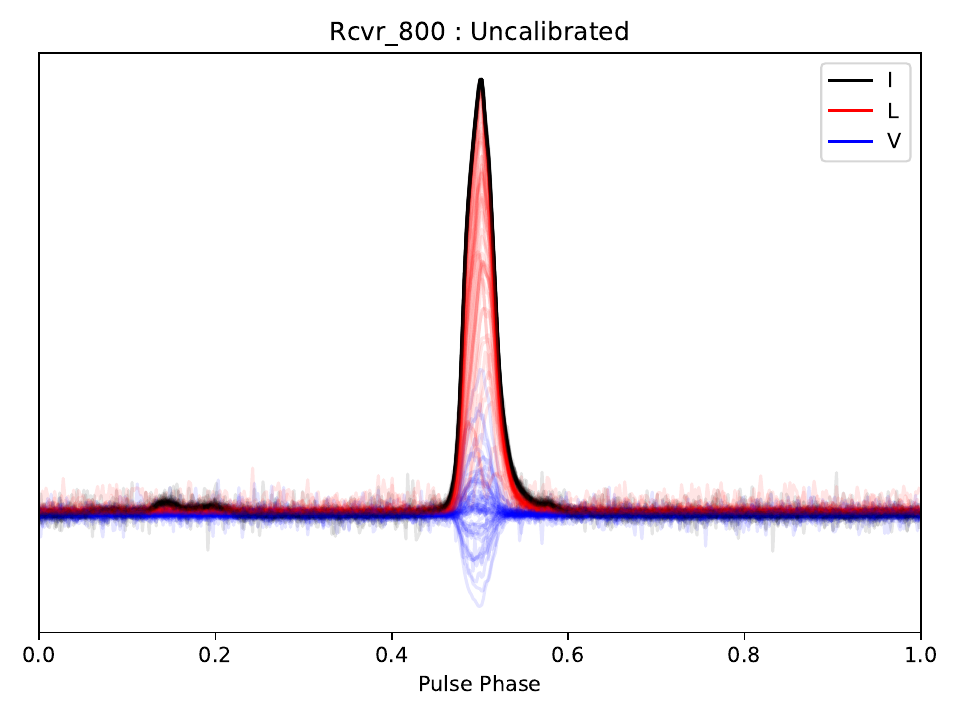}
    \includegraphics[width=0.4\textwidth]{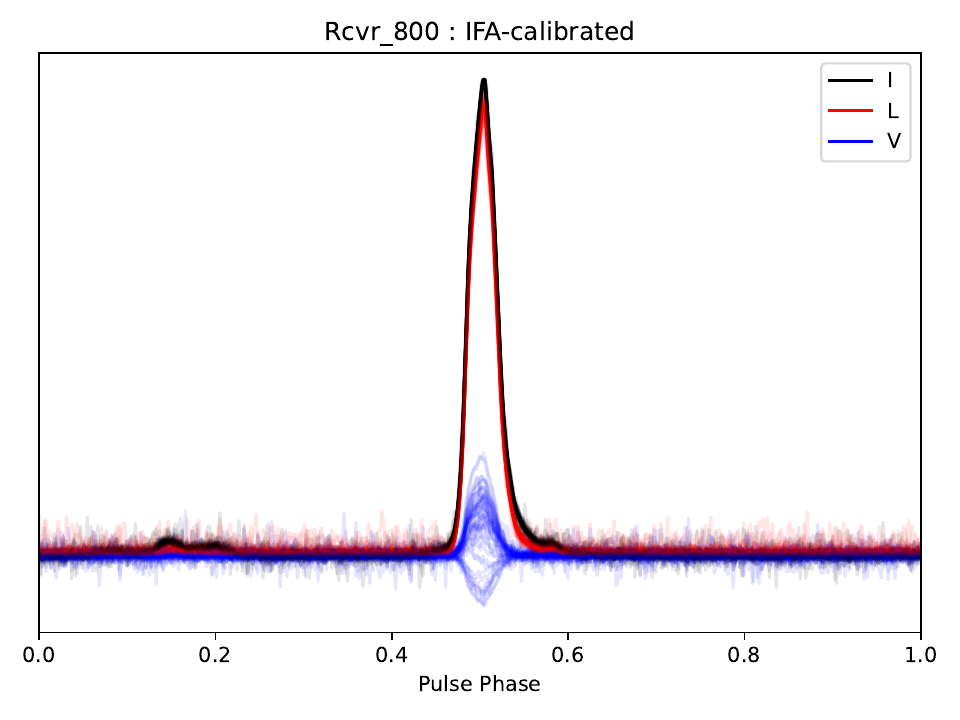}\\
    \includegraphics[width=0.4\textwidth]{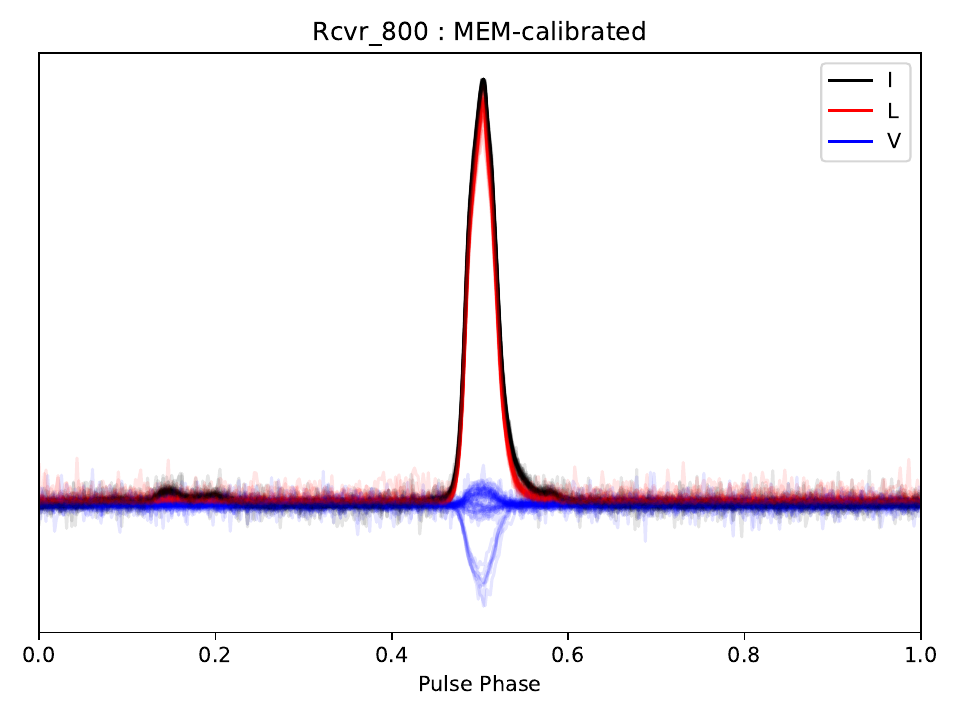}
    \includegraphics[width=0.4\textwidth]{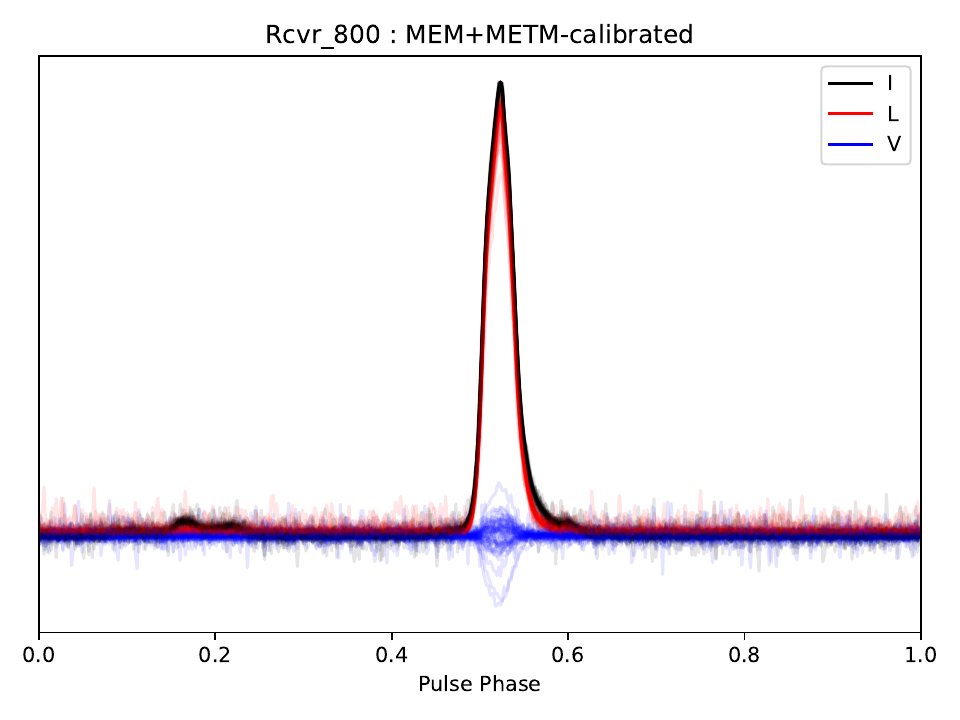}\\
    \includegraphics[width=0.4\textwidth]{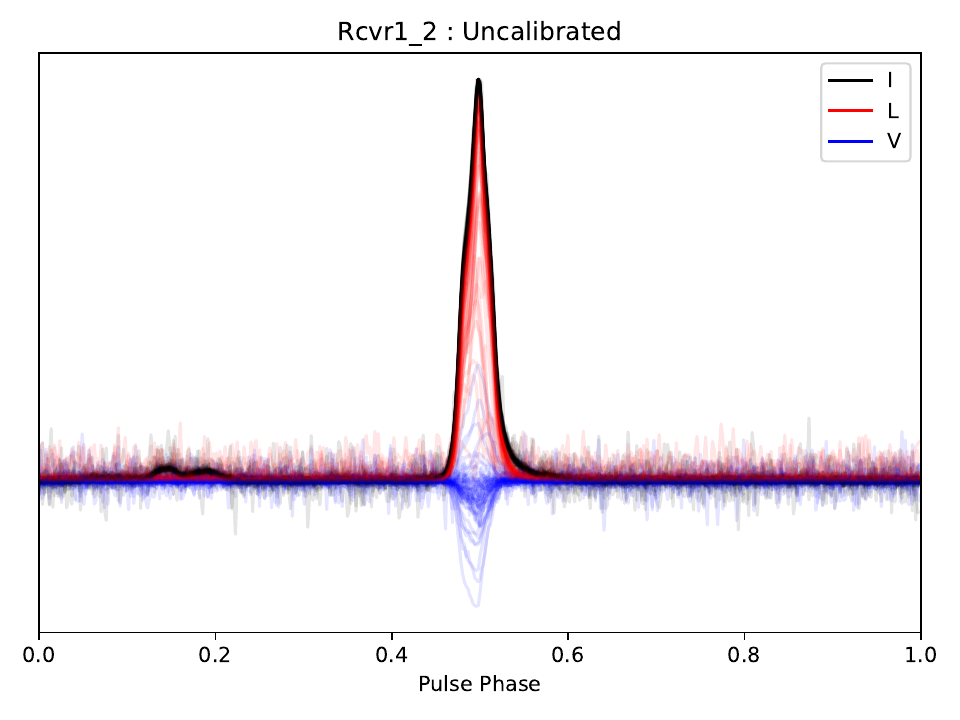}
    \includegraphics[width=0.4\textwidth]{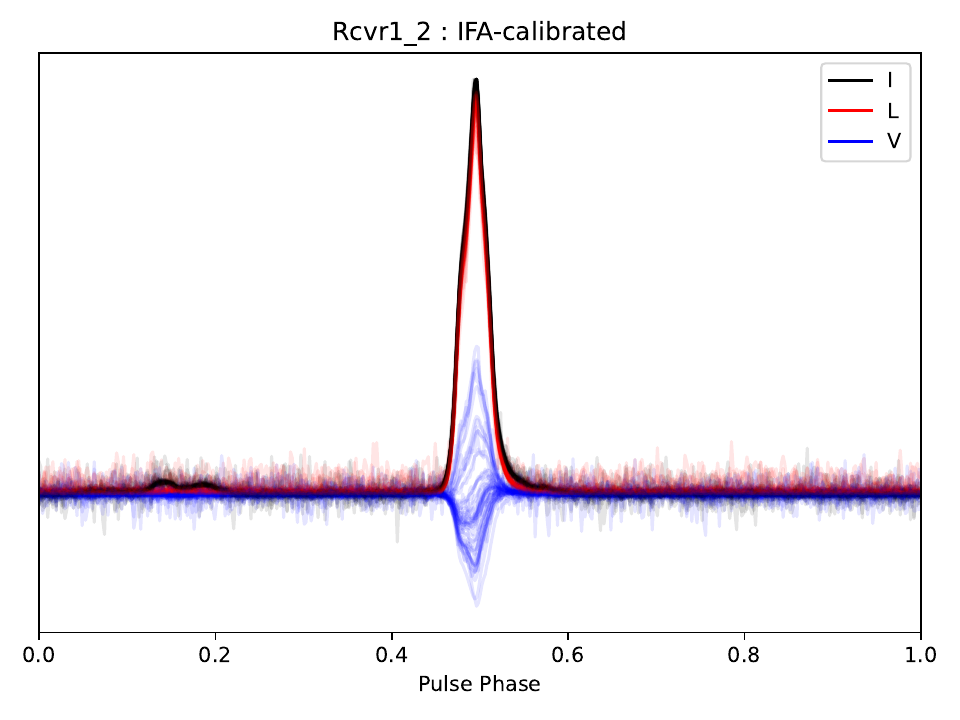}\\
    \includegraphics[width=0.4\textwidth]{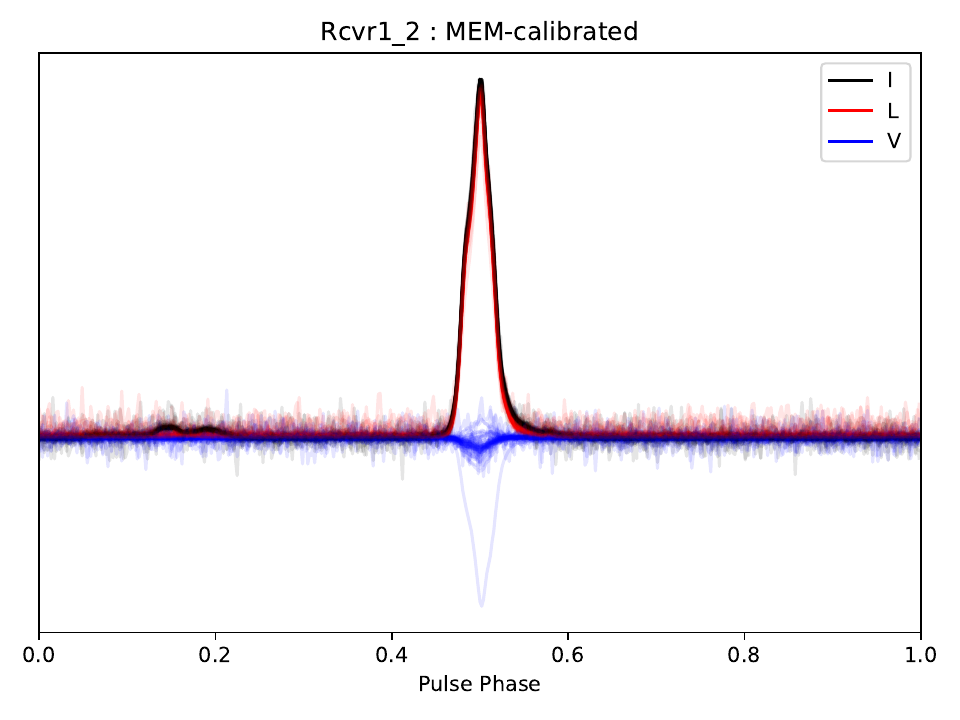}
    \includegraphics[width=0.4\textwidth]{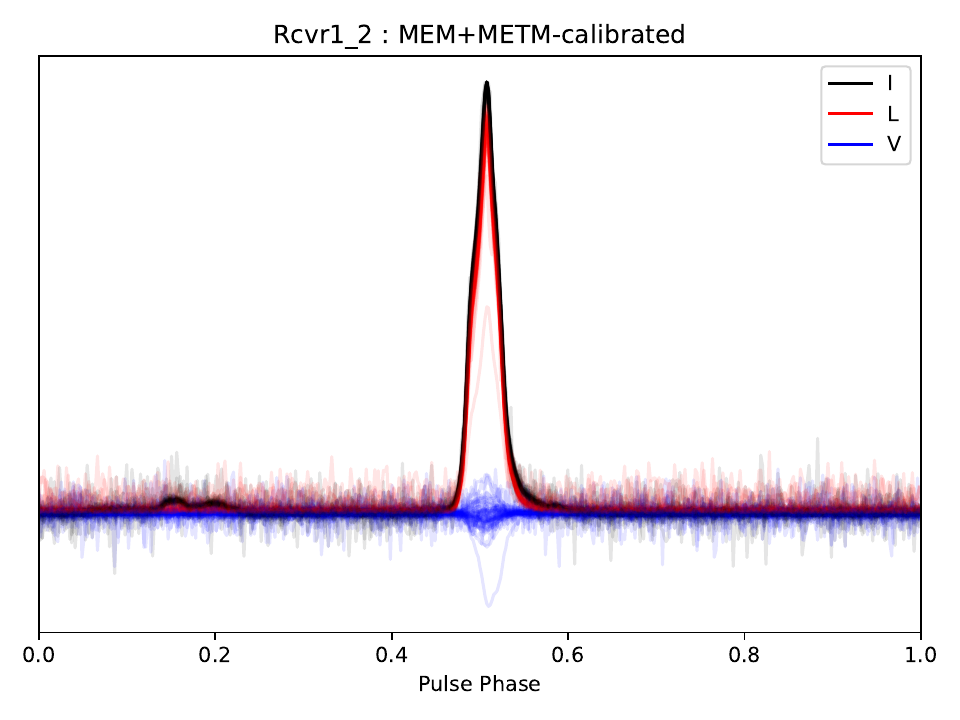}\\
    \caption{Uncalibrated and different calibrated profiles for J1744$-$1134 observed with Rcvr\_800 (800 MHz) and Rcvr1\_2 (1500 MHz) receivers and GUPPI backend system at the GBT.}
    \label{fig:J1744_profs}
\end{figure*}

In this appendix, we present the uncalibrated profiles as well as profiles obtained by different calibration methods for PSRs J1909$-$3744, J1643$-$1224, and J1744$-$1134.
Figure~\ref{fig:J1909_800_profs} shows the uncalibrated and calibrated polarization profiles for J1909$-$3744 observed with Rcvr\_800 and GUPPI backend system at GBT.
Full polarization profiles for J1643$-$1224 and J1744$-$1134, for both Rcvr\_800 and Rcvr1\_2 observation, are shown in Figures~\ref{fig:J1643_profs} and \ref{fig:J1744_profs}, respectively.

\bibliography{bibliography}{}
\bibliographystyle{aasjournal}



\end{document}